\documentclass[10pt,aps,prd,longbibliography,onecolumn,superscriptaddress]{revtex4-2}

\usepackage[T1]{fontenc}
\usepackage[utf8]{inputenc}
\usepackage[english]{babel}

\usepackage{amsfonts,amsbsy,amssymb,amsmath,graphicx,float} 
\usepackage{color}
\usepackage{multirow}

\graphicspath{{figures/}}

\usepackage[linktoc=all,colorlinks=true,citecolor=blue,linkcolor=blue]{hyperref}

\begin{document}

\title{Phase Space Analysis of the Dynamics on a Potential Energy Surface \\ with an Entrance Channel and Two Potential Wells}

\author{Matthaios Katsanikas}
\email{matthaios.katsanikas@bristol.ac.uk}
\affiliation{School of Mathematics, University of Bristol, \\ Fry Building, Woodland Road, Bristol, BS8 1UG, United Kingdom.}

\author{V\'ictor J. Garc\'ia-Garrido}
\email{vjose.garcia@uah.es}
\affiliation{Departamento de F\'isica y Matem\'aticas, Universidad de Alcal\'a, \\ Alcal\'a de Henares, 28871, Spain.}

\author{Makrina Agaoglou}
\email{makrina.agaoglou@bristol.ac.uk}
\affiliation{School of Mathematics, University of Bristol, \\ Fry Building, Woodland Road, Bristol, BS8 1UG, United Kingdom.}

\author{Stephen Wiggins}
\email{s.wiggins@bristol.ac.uk}
\affiliation{School of Mathematics, University of Bristol, \\ Fry Building, Woodland Road, Bristol, BS8 1UG, United Kingdom.}

% \date{\today}

\begin{abstract}

In this paper we unveil the geometrical template of phase space structures that governs transport in a Hamiltonian system described by a potential energy surface with an entrance/exit channel and two wells separated by an index-1 saddle. For the analysis of the nonlinear dynamics mechanisms, we apply the method of Lagrangian descriptors, a trajectory-based scalar diagnostic tool that is capable of providing a detailed phase space tomography of the interplay between the invariant manifolds of the system. Our analysis reveals that the stable and unstable manifolds of the two families of unstable periodic orbits (UPOs) that exist in the regions of the wells are responsible for controlling access to the potential wells of the trajectories that enter the system through the entrance/exit channel. We demonstrate that the heteroclinic and homoclinic connections that arise in the system between the manifolds of the families of UPOs characterize the branching ratio, a relevant quantity used to measure product distributions in chemical reaction dynamics.

\end{abstract}

\maketitle


% \tableofcontents

\section{Introduction}
\label{sec:intro}

One of the biggest challenges in the study of organic chemical reactions is that of providing a sound theoretical understanding of the underlying mechanisms that govern selectivity, i.e. product distributions, in potential energy surfaces (PESs) that display valley-ridge inflection (VRI) points in their topography \cite{rehbein2011,ess2008}. These VRI points, which occur at locations of the PES characterized by two sequential index-1 saddles with no intervening energy minimum (a potential well), are ubiquitous in the chemistry literature and have attracted the attention of both chemists and mathematicians in the past decades \cite{Valtazanos1986,quapp2004,birney2010}. In the vicinity of these points, the intrinsic reaction coordinate bifurcates due to the shape of the PES, and this gives rise to a reaction mechanism known as a two-step-no-intermediate mechanism \cite{singleton2003}. Mathematically, at a VRI point two conditions are met: the Gaussian curvature of the PES is zero, which implies that the Hessian matrix has a zero eigenvalue, and also the gradient of the potential is perpendicular to the eigenvector corresponding to the zero eigenvalue. Geometrically, this means that the landscape of the PES in the neighborhood of the VRI changes its shape from a valley to a ridge.

\begin{figure}[htbp]
	\begin{center}
		\includegraphics[scale=0.28]{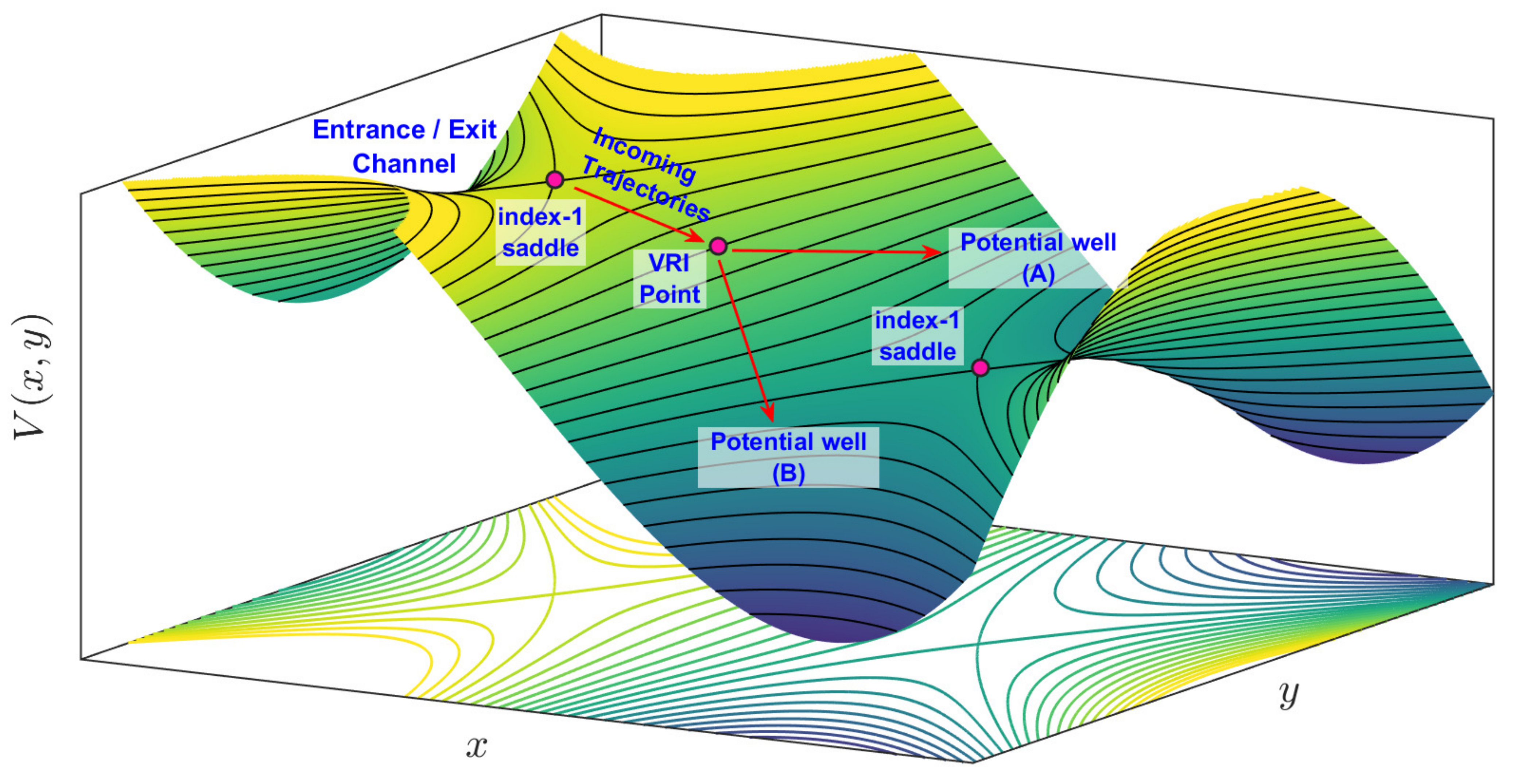}
	\end{center}
	\caption{Potential energy landscape for a physical system exhibiting post-transition state bifurcation.}
	\label{VRI_geometry}
\end{figure}

The goal of this paper is to study reaction dynamics on a symmetric PES exhibiting post-transition state bifurcation in the vicinity of a valley-ridge inflection point. Our work can be viewed as a continuation and extension of the work in \cite{collins2013}. The model PES that we consider has a high energy index-1 saddle point, and a lower energy index-1 saddle that separates two potential wells. In between the two saddle points there is a VRI point. A schematic geometrical representation of the landscape of such PES in the neighborhood of the VRI point is shown in Fig. \ref{VRI_geometry} and discussed in more detail in Section \ref{sec:sec1}. In the work carried out in \cite{collins2013}, trajectories were initiated on a dividing surface located in the region of the higher energy saddle (the upper index-1) at a fixed total energy slightly above that of the saddle, with a value of momentum such that they approached the region of the lower saddle. In the process of evolution, the trajectories crossed the region of the VRI and entered one of the potential wells. The trajectory based quantity of particular interest in this analysis was the relative number of trajectories entering each well, i.e. the {\em branching ratio}.  The nature of the branching ratio is determined by the {\em selectivity}, as it is referred to in the chemistry literature. The PES considered in \cite{collins2013} was not symmetric in the sense that the two wells had different depths. Moreover, parameters in the PES could be varied in a way that slightly changed the location of the wells. It was observed that the branching ratio was sensitive to this change of parameters. 

It was argued in \cite{collins2013} that an understanding of the dynamics underlying the branching ratio in such a PES and the mechanism underlying the nature of the selectivity required a phase space analysis, i.e. an analysis that explicitly considered the influence of the momentum of trajectories. This is particularly true when the underlying dynamics is nonstatistical, and the number of observations of organic reactions exhibiting nonstatistical behaviour is increasing yearly. An overview of many of these, as well as a guide to several reviews discussing the subject, is given in \cite{collins2013}. The importance of understanding the dynamical mechanisms of selectivity was underlined by noting that control of selectivity is of essential importance for synthesis, especially if existing models used for analyzing the problem are incomplete or inapplicable. 

In this paper we consider a symmetric version of the potential energy surface studied in \cite{collins2013}. This symmetric version allows us to uncover the dynamical origin of the mechanism underlying selectivity in an unambiguous manner. This, again, highlights the necessity of a phase space perspective for understanding the dynamics. The phase space approach for analyzing chemical dynamics is reviewed in \cite{Agaoglou2019}, which contains an extensive guide to the literature. In this work, we will use three particular techniques of phase space nonlinear dynamics - Poincar\'{e} maps, Lagrangian descriptors and lobe dynamics.

Poincar\'e maps \cite{henon1982,contopoulos2002} allow us to reveal and analyze the Kolmogorov-Arnold-Moser (KAM) tori associated to the regular behavior displayed by the system. KAM tori are significant because they lead to trapping of trajectories. However, Poincar\'e maps provides an incomplete dynamical picture, since the regions where trajectories are chaotic appear in the Poincar\'{e} surfaces of section (PSOS) as a chaotic sea of random points, which are extremely difficult to interpret and completely obscure the intricate interactions (tangles) between the invariant stable and unstable manifolds. For this reason, we complement the analysis with Lagrangian descriptors (LDs) \cite{madrid2009,mancho2013lagrangian,lopesino2017}, which are a scalar trajectory diagnostic technique with the capability of revealing the stable and unstable manifolds and their intricate interactions through the formation of lobes. The resulting lobe dynamics provide us with a way to quantify phase space \cite{rkw,wigginschaotictransport} and form the basis of our understanding of selectivity.

The method of Lagrangian descriptors is a nonlinear dynamics technique that was first introduced a decade ago to analyze Lagrangian transport and mixing processes in Geophysical flows \cite{madrid2009,mendoza2010}. The first definition of LDs relied on the computation of the arclength of trajectories of initial conditions as they evolve forward and backward in time \cite{mendoza2010,mancho2013lagrangian}. Since its proposal, this methodology has found a myriad of applications in different scientific areas. For instance, in the context of Geophysics, it has been used in Oceanography to plan transoceanic autonomous underwater vehicle missions by taking advantage of the underlying dynamical structure of ocean currents \cite{ramos2018}. Also, it has been shown to provide relevant information for the effective management of marine oil spills \cite{gg2016}. Recently, this tool has also received recognition in the field of Chemistry, for instance in Transition State Theory \cite{craven2015lagrangian,craven2016deconstructing,craven2017lagrangian,revuelta2019unveiling}, where the computation of chemical reaction rates relies on the knowledge of the phase space structures that separate reactants from products. Other applications of this tool to chemical problems include the analysis of isomerization reactions \cite{GG2020b} and roaming \cite{krajnak2019,gonzalez2020}, the study of the influence of bifurcations on the manifolds that control chemical reactions \cite{GG2020a}, and also the explanation of the dynamical matching mechanism in terms of the existence of heteroclinic connections in a Hamiltonian system defined by Caldera type PES \cite{katsanikas2020a}.

The work discussed in this paper is an extended and more detailed study of the analysis we started in \cite{Agaoglou2020}. The contents of this paper are organized as follows. In Section \ref{sec:sec1} we describe the fundamental landscape characteristics of the PES which defines the two degrees-of-freedom (DoF) Hamiltonian model used in this work for the analysis of the phase space transport processes, including the selectivity mechanism. Section \ref{sec:sec3} is devoted to providing a detailed description of the results obtained from the analysis of the nonlinear dynamics of the system. Finally, Section  \ref{sec:conc} summarizes the conclusions of this paper. The reader can find in Appendix \ref{sec:appA} a brief explanation on the method of Lagrangian descriptors, and how this technique can be applied to easily reveal the geometrical template of invariant manifolds and their intricate heteroclinic and homoclinic connections in the high-dimensional phase space of Hamiltonian systems.

\section{The Hamiltonian Model for the Branching Mechanism}
\label{sec:sec1}

In this section we present the fundamental characteristics of the Hamiltonian model with two DoF that we study in this work. The potential energy surface that defines our Hamiltonian model is inspired in the analysis carried out in \cite{collins2013}, where the influence of valley-ridge inflection points of the PES on the selectivity (branching) mechanism observed in many organic chemical reactions is addressed. The topography of the PES introduced in \cite{collins2013} has an entrance/exit channel characterized by an index-1 saddle and two potential wells separated by an energy barrier determined by another index-1 saddle. The PES also has a VRI point between both saddles, where the intrinsic reaction coordinate of the system, i.e. the minimum energy path, bifurcates (or branches). It is thought in the Chemistry literature that this geometrical phenomenon observed in the vicinity of VRI points might play a relevant role in the determination of selectivity mechanisms, ``guiding'' the  trajectories that enter the system through the channel towards any of the potential wells.

\begin{figure}[htbp]
	\begin{center}
	\includegraphics[scale=0.35]{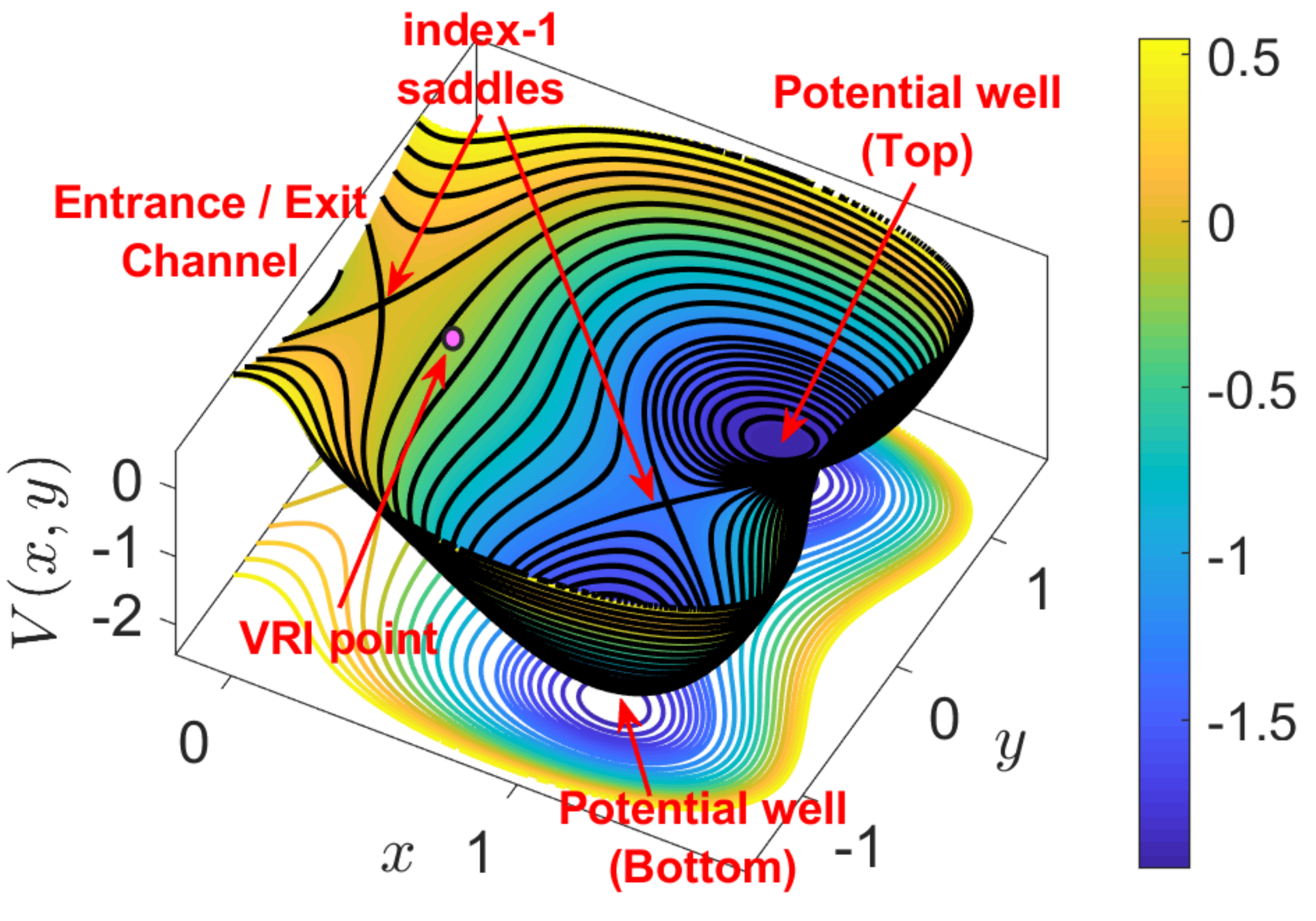}
	\includegraphics[scale=0.34]{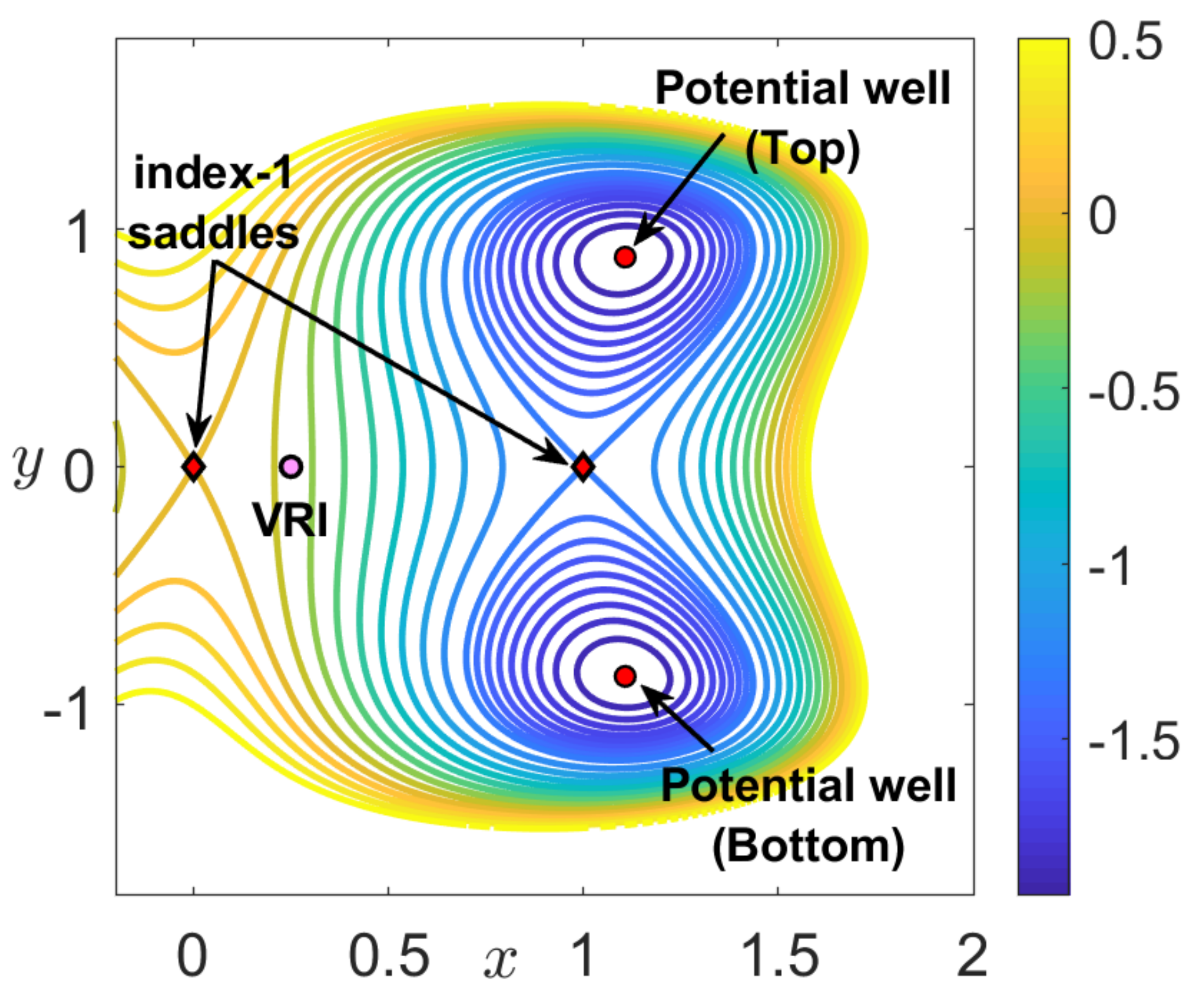}
	\end{center}
	\caption{Potential energy surface given in Eq. \eqref{pes_modelVRI}. We have marked the relevant dynamical features of the energy landscape.}
	\label{PES_plot}
\end{figure}

In this paper we use a simplified version of the PES from the one discussed in \cite{collins2013}, where we assume that the energy landscape is symmetric with respect to the $x$-axis. Our Hamiltonian model has the classical structure of kinetic plus potential energy in the form:
\begin{equation}
H(x,y,p_x,p_y) = \dfrac{p_x^2}{2 m_x} + \dfrac{p_y^2}{2 m_y} + V(x,y) \;,
\label{hamiltonian}
\end{equation}
where we assume that the mass in each DoF is $m_x = m_y = 1$, and the PES is described by:
\begin{equation}
V(x,y) = \dfrac{8}{3}x^3 - 4x^2 + \dfrac{1}{2} y^2 + x\left(y^4 - 2 y^2\right) \;.
\label{pes_modelVRI}
\end{equation}  
The PES is symmetric with respect to the $y$ coordinate and it has two wells separated by an index-1 saddle located at the point $(1,0)$. We label this saddle as the lower index-1 saddle. Moreover, the energy landscape has an entrance/exit channel determined by an index-1 saddle located at the origin, which we call the upper saddle and has the highest energy among all the critical points on the PES. An illustration of the geometry of the PES described above is included in Fig. \ref{PES_plot}, and the location and energies of all the critical points is summarized in Table \ref{tab:tab1}. Recall that, for a two DoF Hamiltonian, potential wells are local minima of the PES and the Hessian matrix evaluated at these critical points has two positive eigenvalues. On the other hand, index-1 saddles of the PES are critical points of saddle type, that is, the Hessian matrix evaluated at them has a positive and a negative eigenvalue. Regarding the VRI point, although it is not a critical point of the PES, we point out that it lies on the $x$ axis at the location $(1/4,0)$, and its energy is given by $V_{I} = V(1/4,0) = - 5/24$. 

The evolution of the Hamiltonian system in Eq. \eqref{hamiltonian} takes place in a 4-dimensional phase space, and is determined by Hamilton's equations of motion:
\begin{equation}
\begin{cases}
\dot{x} = \dfrac{\partial H}{\partial p_x} = p_x \\[.5cm]
\dot{y} = \dfrac{\partial H}{\partial p_y} = p_y  \\[.5cm]
\dot{p}_x = -\dfrac{\partial H}{\partial x} = 8 x \left(1 - x\right) + y^2\left(2 - y^2 \right) \\[.5cm]
\dot{p}_y = -\dfrac{\partial H}{\partial y} = y \left[4 x \left(1 - y^2\right) - 1\right]
\end{cases}
\;.
\label{ham_eqs}
\end{equation}
Since energy is conserved, dynamics of trajectories is constrained to a three-dimensional energy hypersurface. It is important to remark here that, due to the symmetry of the PES in Eq. \eqref{pes_modelVRI} with respect to the $x$ axis, which is a consequence of the PES being an even function of the $y$ coordinate, i.e. $V(x,y) = V(x,-y)$, the structures in the phase space of the system are symmetric under a $180^{\circ}$ rotation about the origin in the $y$-$p_y$ plane. This symmetry plays, as we will show, a fundamental role for the explanation that the branching ratio for systems with this type of symmetric PES is unity. This means that if we consider all the trajectories that enter the system through the channel of the upper index-1 saddle and visit either well for the first time along their evolution, half of them visit the top well and the other half go to the bottom well.

\begin{table}[htbp]	
	\begin{tabular}{| l | c | c | c | c |}
		\hline
		Critical point \hspace{1cm} & \hspace{0.6cm} x \hspace{0.6cm} & \hspace{.6cm} y \hspace{.6cm} & \hspace{.2cm} \text{Potential Energy} $(V)$ \hspace{.2cm} & \hspace{.6cm} \text{Stability} \hspace{.6cm} \\
		\hline\hline
		index-1 saddle (Upper) \hspace{.5cm} & 0 & 0 & 0 & saddle $\times$ center \\
		\hline
		index-1 saddle (Lower) \hspace{.5cm} & 1 & 0 & -4/3 & saddle $\times$ center \\
		\hline
		Potential Well (Top) \hspace{.5cm} & 1.107146 & 0.879883 & -1.94773 & center \\
		\hline
		Potential Well (Bottom) \hspace{.5cm} & 1.107146 & -0.879883 & -1.94773 & center  \\
		\hline
		\end{tabular} 
		\caption{Location of the critical points of the potential energy surface.} 
	\label{tab:tab1} 
\end{table}

\section{Results}
\label{sec:sec3}

In this section we describe the results obtained in our analysis of the phase space transport and trapping mechanisms that take place in the model Hamiltonian system. We divide this discussion into three different cases, depending on the energy levels of the system:
\begin{enumerate}
\item[{\bf A.}] \underline{{\bf First case:}} The energy is below that of the lower index-1 saddle and above that of the potential wells.
\item[{\bf B.}] \underline{{\bf Second case:}} The energy is below that of the upper index-1 saddle and above that of the lower index-1 saddle.
\item[{\bf C.}] \underline{{\bf Third case:}} The energy is above that of the upper index-1 saddle.
\end{enumerate}
The study of the dynamics is carried out by applying LDs in order to determine the geometry of the invariant stable and unstable manifolds of the unstable periodic orbits present in the system, and we also make use of the method of Poincar\'{e} maps for the analysis of the KAM tori that characterize the regular dynamics. All this analysis is done in the Poincar\'e surfaces of section (PSOS) given below:
\begin{equation}
\begin{split}
\Sigma_1(H_0) & = \left\{ \left(x,y,p_x,p_y\right) \in \mathbb{R}^4 \;\Big| \; x = 0.05 \; , \; p_{x}\left(x,y,p_y;H_0\right) > 0 \right\} \\[.1cm] 
\Sigma_2(H_0) & = \left\{ \left(x,y,p_x,p_y\right) \in \mathbb{R}^4 \;\Big| \; x = 1 \; , \; p_{x}\left(x,y,p_y;H_0\right) > 0 \right\} \\[.1cm]
\Sigma_3(H_0) & = \left\{ \left(x,y,p_x,p_y\right) \in \mathbb{R}^4 \;\Big|\; y = 0 \; , \; p_{y}\left(x,y,p_x;H_0\right) > 0 \right\}\\[.1cm]
\Sigma_4(H_0) & = \left\{ \left(x,y,p_x,p_y\right) \in \mathbb{R}^4 \;\Big| \; x = x_{well} \; , \; p_{x}\left(x,y,p_y;H_0\right) > 0 \right\}
\end{split}
\label{PSOS}
\end{equation}
where $x_{well}$ is the $x$ coordinate of the potential wells, as shown in Table \ref{tab:tab1}. We briefly explain next the reasons for choosing the PSOS defined above. The first section, $\Sigma_1(H_0)$, is taken at the entrance/exit channel and is used to understand the dynamical behavior and fate of the trajectories that enter the system coming from infinity through the phase space bottleneck associated to the upper index-1 saddle. On the other hand, sections $\Sigma_2(H_0)$ and $\Sigma_3(H_0)$ are used in order to analyze the regular dynamics governed by the KAM tori in the system and address well to well transport. Finally, the purpose of $\Sigma_4(H_0)$ is to illustrate the bifurcation of the family of KAM tori corresponding to the potential wells at a certain value of the energy of the system.

\subsection{First case}

The first case corresponds to an energy range for the system that goes from the energy of the potential wells (stable equilibrium points of Hamilton's equations) to that of the lower index-1 saddle that separates both wells. For this energy regime, the well regions are not connected in the phase space, since the bottleneck associated to the lower index-1 saddle that sits between them is closed. Therefore, trajectories are forbidden to evolve from well to well, and hence, they are forever trapped in either well. Moreover, in this situation, each stable equilibrium point has at least two families of periodic orbits according to the Lyapunov subcenter theorem \cite{weinstein1973,moser1976,rabinowitz1982}. In Fig. \ref{well_bif_negEn} A) we observe the KAM tori \cite{kolmogorov1954,moser1962,arnold1963} associated to the family of stable periodic orbits that exists in the region of the top well. We do so by computing the Poincar\'e map in the section $\Sigma_{4}(H_{0})$ for the energy value $H_0 = -1.4$. Notice that, although we only show in the figure the region of the top well, the same structures would appear in the bottom well due to the symmetry property of the PES with respect to the $y$ coordinate that we discussed in the previous section. This will induce a $180^{\circ}$ rotational symmetry about the origin for the coordinates $y$ and its canonically conjugate momentum $p_y$.

\subsection{Second case}

We turn our attention next to the case where the energy of the system is above that of the lower index-1 saddle between both wells, but below that of the upper index-1 saddle. In this energy interval, the entrance/exit channel associated to the upper index-1 is still closed so that no trajectories can escape the system. Moreover, for these energy values, a phase space bottleneck connects both wells and therefore many trajectories can move back and forth between them, crossing along their evolution the index-1 saddle region. This type of trajectories are known as 'reactive', as opposed to those that stay forever in either well, which are labelled as 'trapped' or 'non-reactive'.

In this setup we analyze the phase space structures that determine transport in the phase space of the system from one well to the other. According to the Lyapunov subcenter theorem \cite{weinstein1973,moser1976,rabinowitz1982}, we know that there exists at least one family of unstable periodic orbits associated to the index-1 saddle separating the wells. This UPO, which has the topology of a circle, characterizes the bottleneck region that reactive trajectories have to cross in order to evolve between wells. Attached to the UPO we have two-dimensional stable and unstable manifolds, known in the literature as spherical cylinders, with the form of tubes. These structures are responsible for controlling transport across the index-1 region in the phase space. Initial conditions lying inside the stable/unstable manifold tube will cross the bottleneck in forward/backward time respectively, and the fate of those trajectories outside the tubes is to remain trapped forever in the well region where they started.   

\begin{figure}[htbp]
	\begin{center}
		A)\includegraphics[scale=0.28]{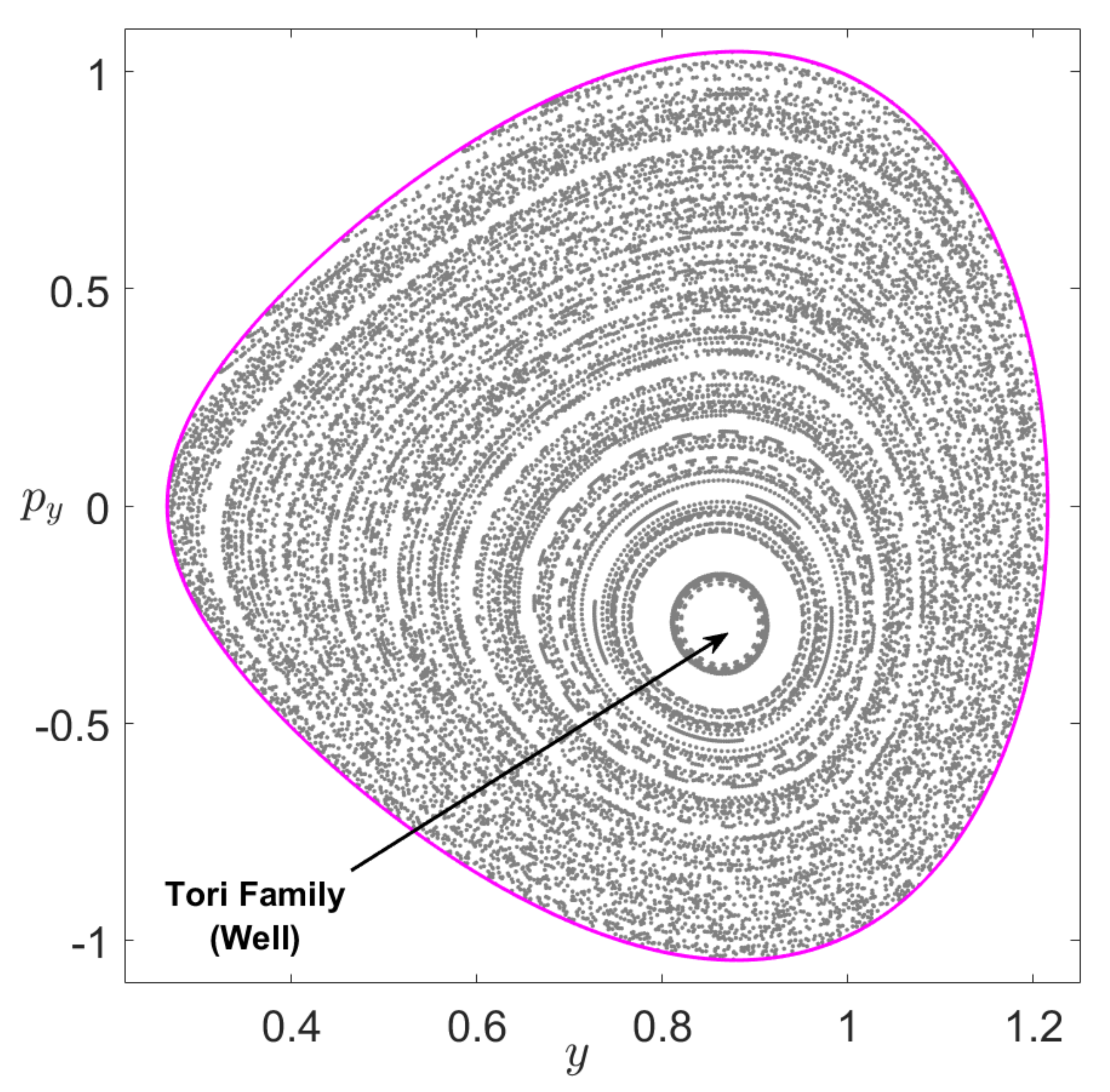}
		B)\includegraphics[scale=0.28]{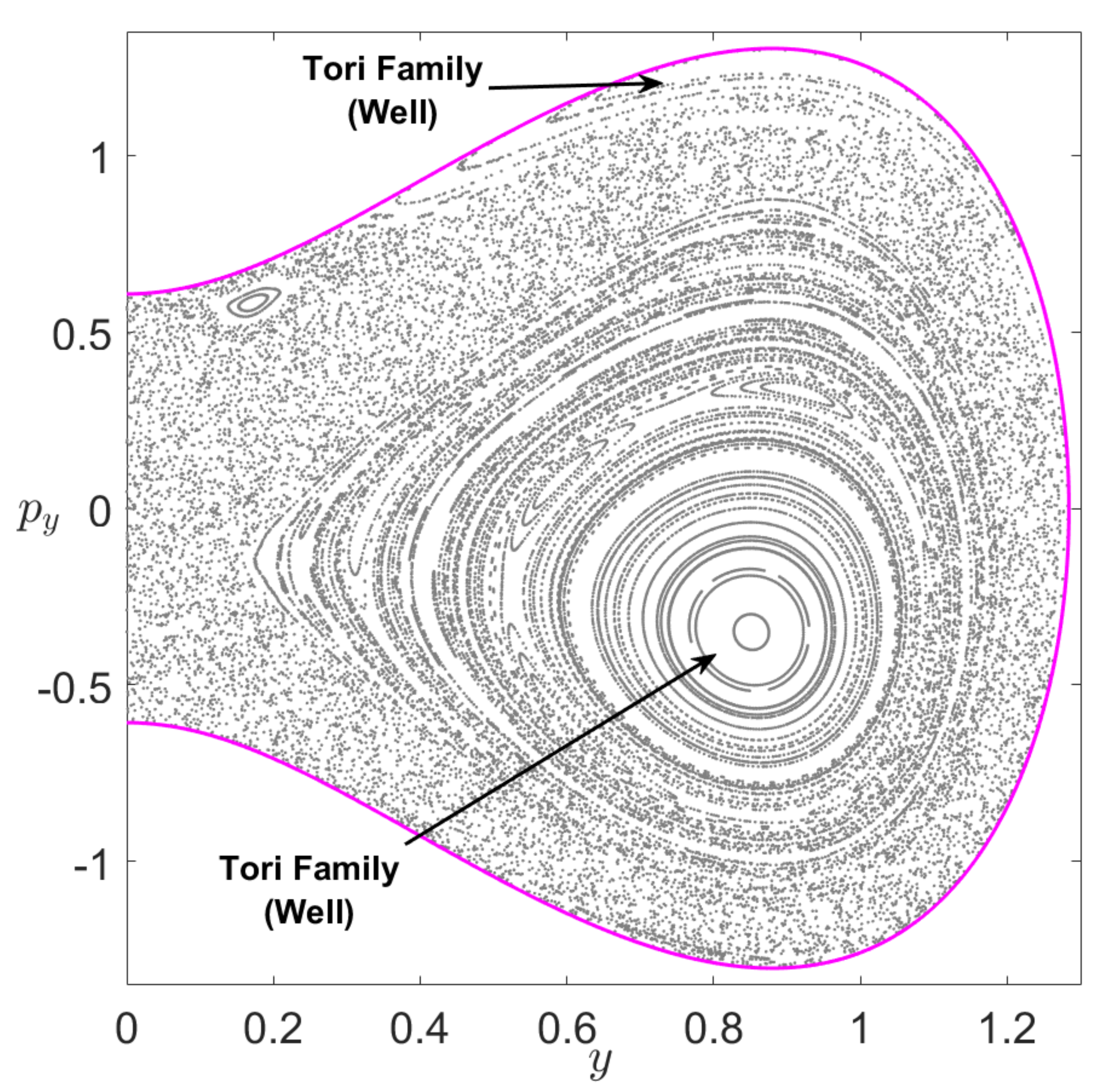}
	\end{center}
	\caption{Poincar\'e map calculated on the surface of section $\Sigma_4(H_0)$ for the system's energy: A) $H_0 = -1.4$; B) $H_0 = -1.1$}
	\label{well_bif_negEn}
\end{figure}

We probe the system dynamics by taking first an energy $H_0 = -1.1$. If we compute a Poincar\'e map in the section $\Sigma_4(H_0)$ we can see in Fig. \ref{well_bif_negEn} B) that the phase space region where chaotic dynamics occurs has grown in size with respect to what we observed in Fig. \ref{well_bif_negEn} A). Moreover, notice the appearance of another stable periodic orbit with KAM tori (islands of regularity) around it at the top of the energy boundary displayed in magenta. This periodic orbit is associated to the wells and belongs to a family with period 2, that is, it has two branches. One branch is close to the top well, and the other, to the bottom well. If we select an initial condition in one of these tori, the resulting trajectory will display regular quasiperiodic motion and it will move back and forth between both wells of the PES. On the other hand, a trajectory that starts from the regular region located at the lower-right part of Fig. \ref{well_bif_negEn} will be trapped in the well. Other trajectories initialized in the chaotic sea of Fig. \ref{well_bif_negEn} will cross the index-1 saddle and move from well to well.

\begin{figure}[htbp]
	\begin{center}
	    A)\includegraphics[scale=0.3]{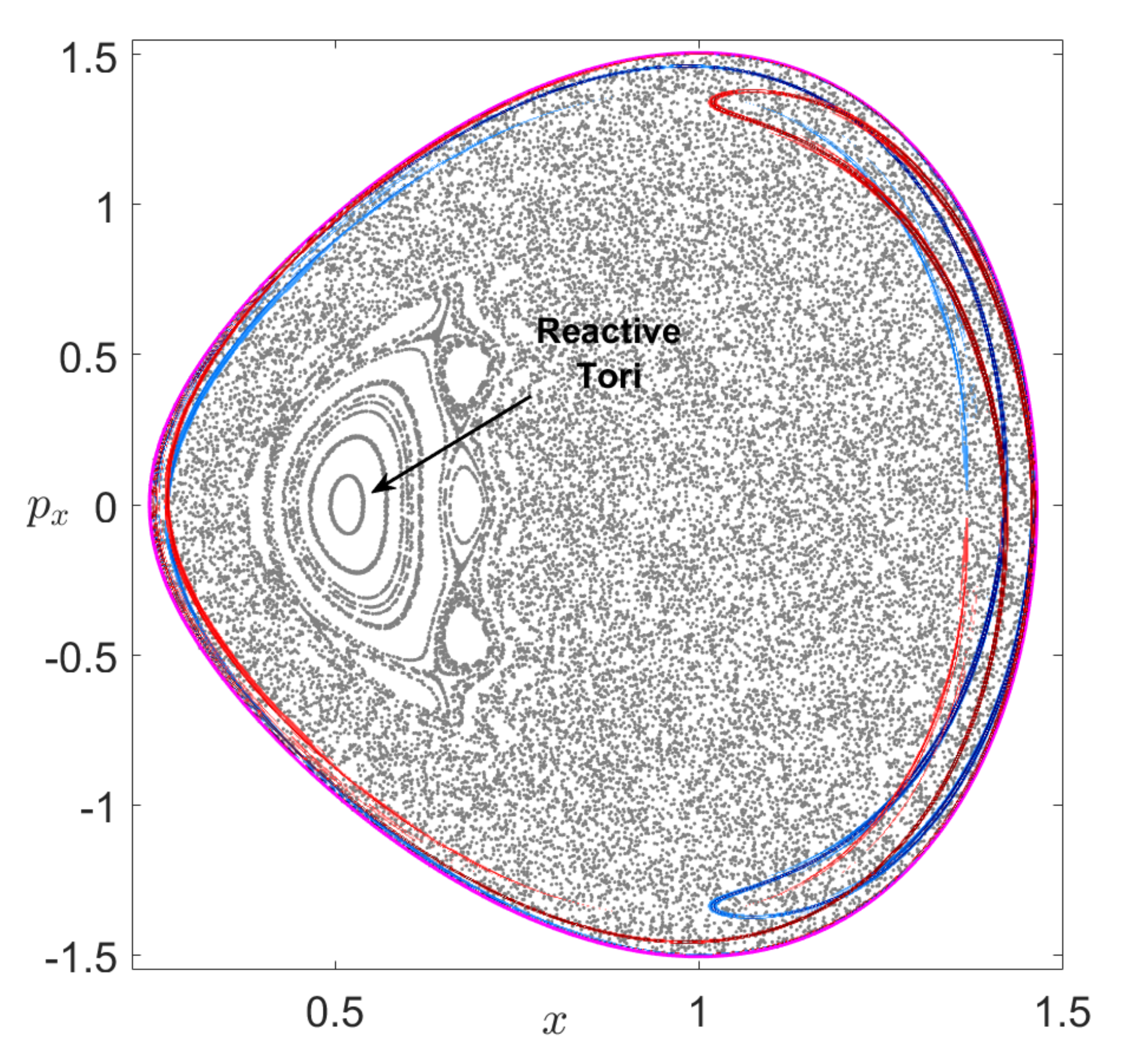}
		B)\includegraphics[scale=0.3]{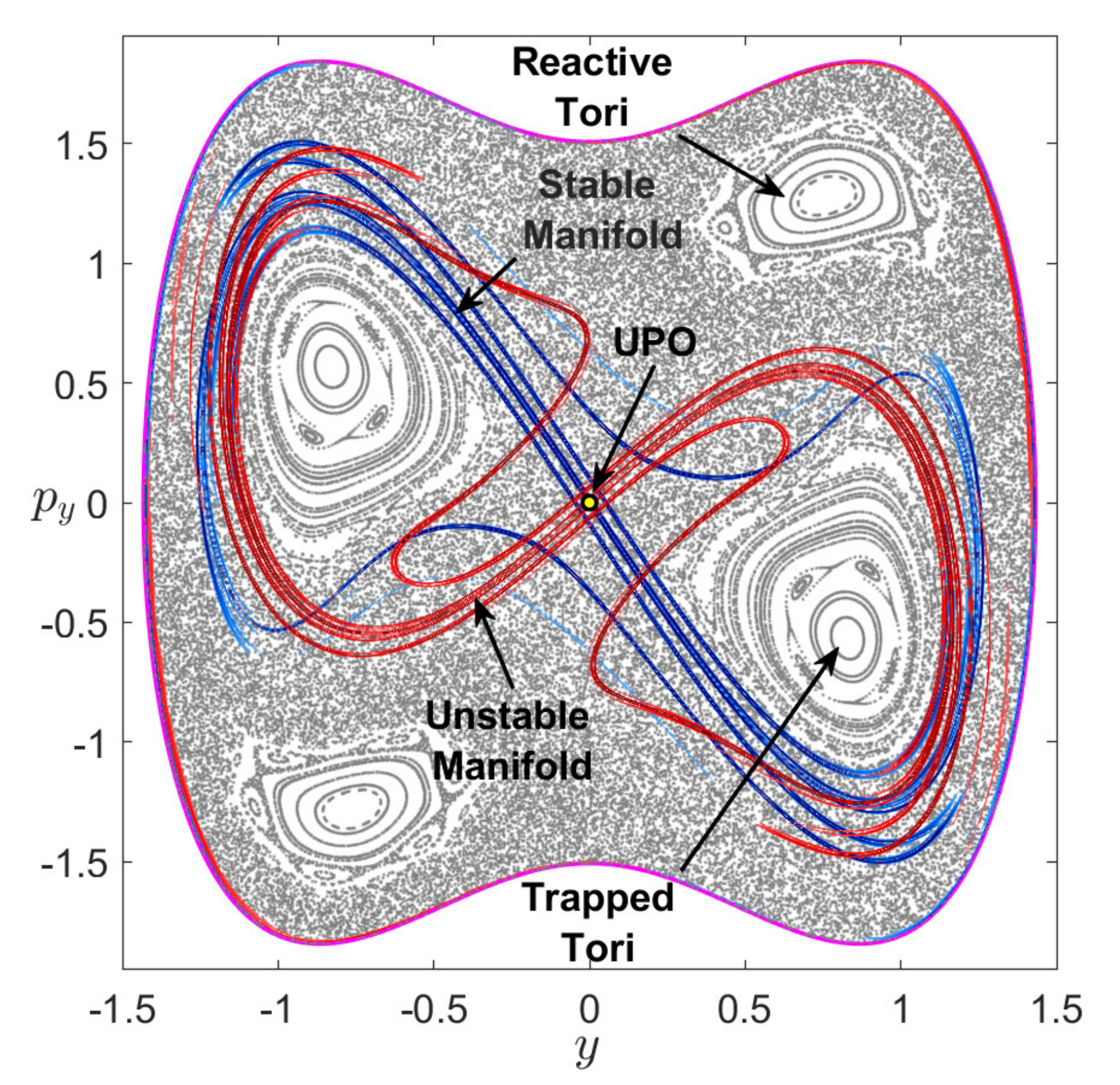}
		C)\includegraphics[scale=0.3]{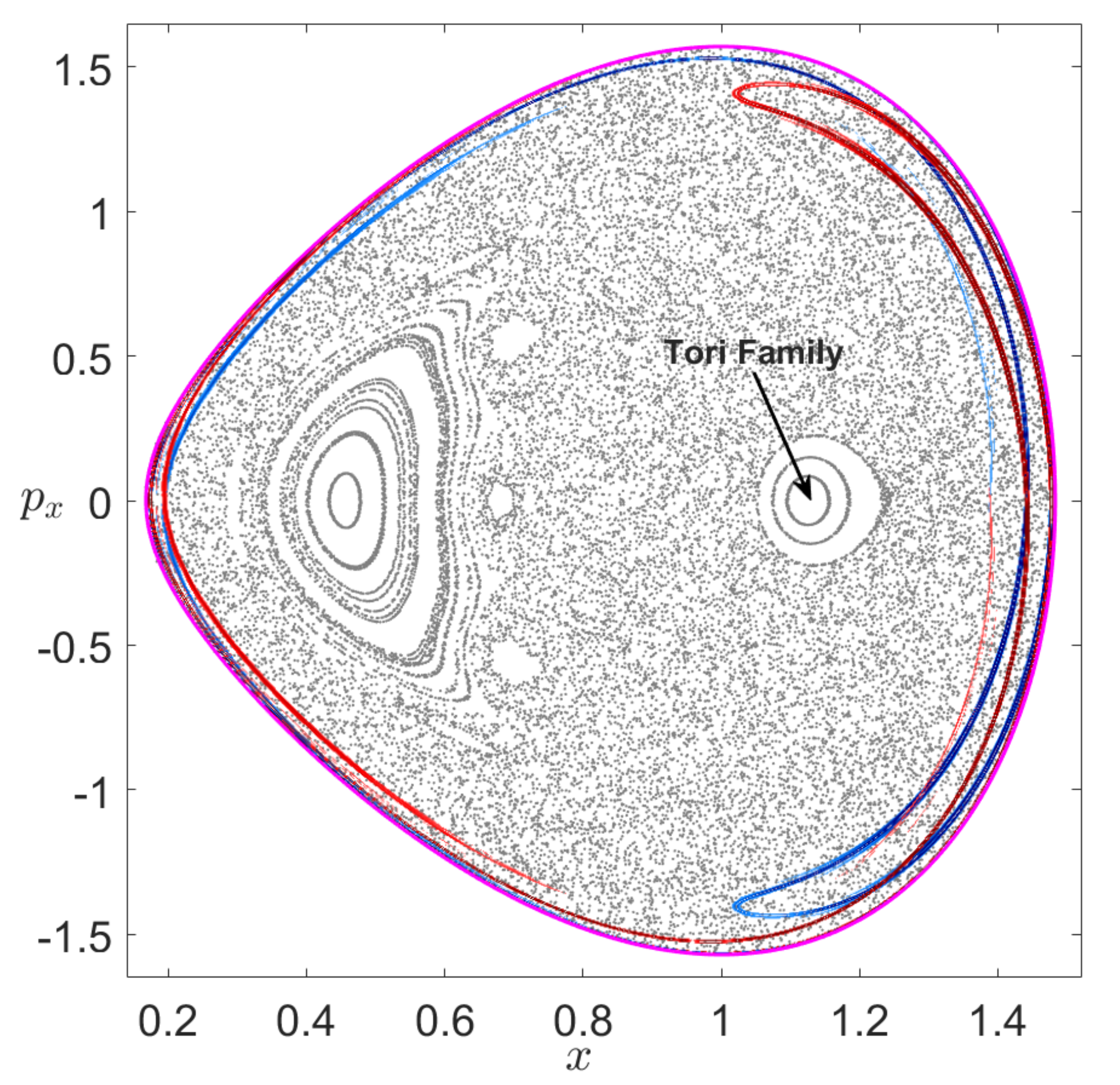}
	    D)\includegraphics[scale=0.3]{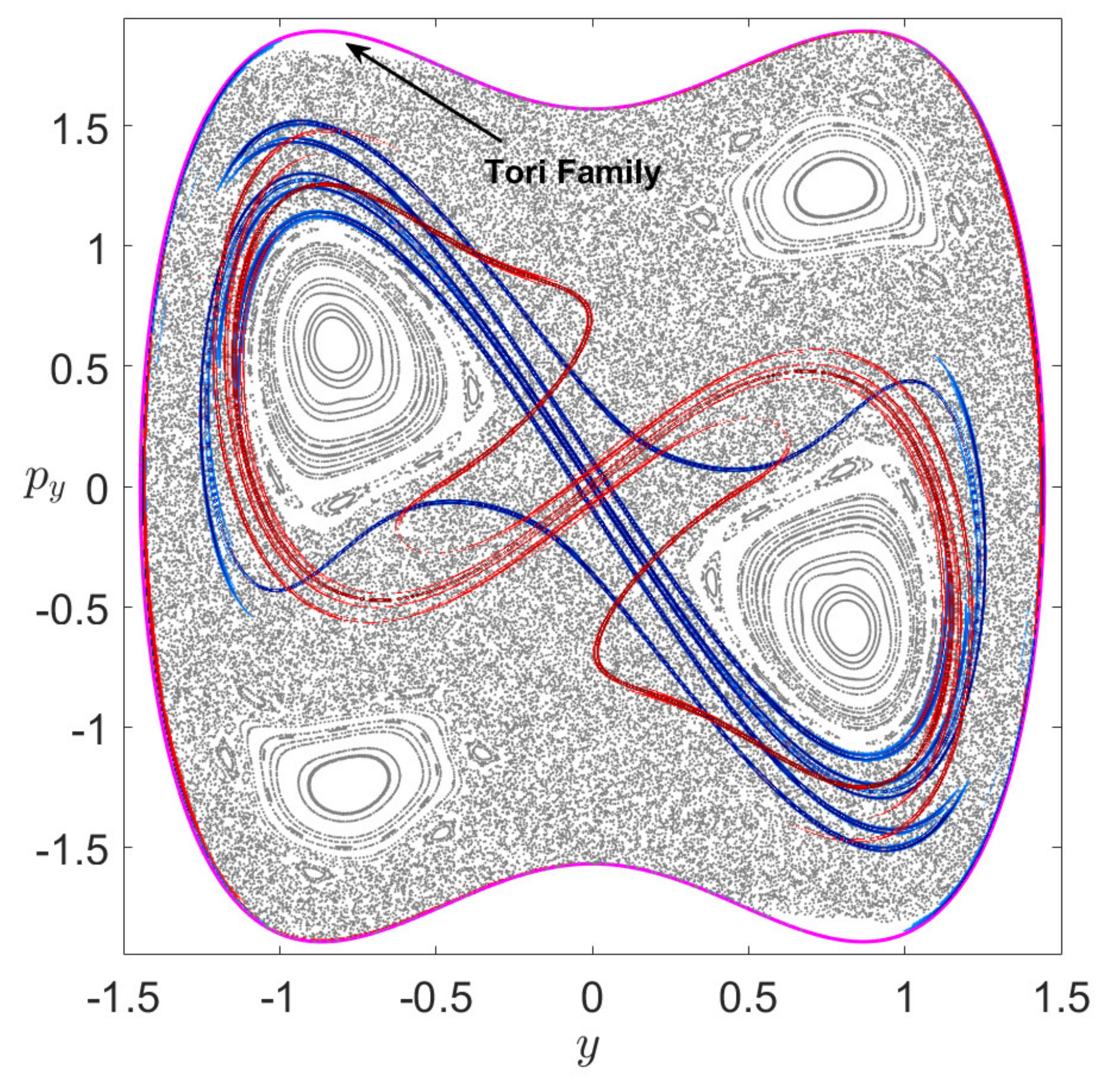}
	\end{center}
	\caption{Poincar\'e map superimposed with the stable (blue) and unstable (red) manifolds extracted from Lagrangian descriptors using $\tau = 8$. A) and C) correspond to the surface of section $\Sigma_3(H_0)$; B) and D) are for the surface of section $\Sigma_2(H_0)$; The energy of the system is $H_0 = -0.2$ and $H_0 = -0.1$ for the panels in the first and second rows respectively. The energy boundary is depicted as a magenta curve.}
	\label{en_bs_sos_ld}
\end{figure}

In order to explore the dynamics of the system further, we consider also the Poincar\'{e} sections $\Sigma _{2}(H_{0})$ and $\Sigma _{3}(H_{0})$, which are very useful for the analysis of the trapping of trajectories in one of the well regions of the PES, and also to study the transport of trajectories from the region of the index-1 saddle that separates both wells to the region of either well. Firstly, we take a look at the phase space structures and trapping close to the family of UPOs associated to the index-1 saddle. We do so for an energy of the system $H_{0} = -0.2$. In Fig. \ref{en_bs_sos_ld} A) we display the Poincar\'e map computed on the surface of section $\Sigma_{3}(H_{0})$, superimposed with the stable (blue) and unstable (red) manifolds extracted from the gradient of the scalar field generated by LDs. Around the families of stable periodic orbits of the wells, we see invariant curves that represent the KAM tori \cite{kolmogorov1954,arnold1963,moser1962}. Trajectories lying on these tori are reactive trajectories, since they are located inside the homoclinic lobes formed by the stable and unstable manifolds of the UPO associated to the index-1 saddle. Therefore, they move from well to well crossing the phase space bottleneck region around the index-1 saddle, and for this reason we label these invariant curves as 'reactive tori'.

Now we want to explore in detail the transport mechanism of trajectories between the two wells. We compute the Poincar\'e map and LDs on the surface of section $\Sigma_{2}(H_{0})$ and we show the results in Fig. \ref{en_bs_sos_ld} B). In the upper right and lower left corners of this figure we can still see the islands of regularity around the stable periodic orbit with period 2 associated to the family of the wells. These 'reactive tori' coincide with those we revealed in panel A. Notice also that the we can clearly observe the symmetry in the phase space structures of the system, which is induced by the symmetry of the PES with respect to the $y$ variable. On the other hand, those tori that we label as 'trapped tori' correspond to trajectories that display regular motion and that do not cross the index-1 saddle region. Therefore, these KAM tori cannot appear in the surface of section $\Sigma_{3}(H_{0})$ (that corresponds to $y = 0$) shown in panel A. In Fig. \ref{en_bs_sos_ld} B) we can also observe the stable and unstable manifolds of the UPO associated to the index-1 saddle that separates both wells. Transport between both wells is explained by the homoclinic intersections between the stable and unstable manifolds of the UPO of the index-1 saddle. It is important to remark the $180^{\circ}$ rotational  symmetry displayed by the manifolds about the origin, which is a consequence of the symmetry of the PES. This implies a symmetric transport mechanism in the system. What we mean by this is that trajectories that are transported from well to well are guided equally from the unstable manifolds of the UPO of the index-1 saddle to each region of two wells ($50\%$ of the trajectories to the one well and $50\%$ to the other). 

We increase next the energy of the system to $H_{0} = -0.1$ in order to study the  trapping and transport mechanisms for higher values of the energy. We present these results in Fig. \ref{en_bs_sos_ld} C) and D) for the sections $\Sigma _{3}(H_{0})$ and $\Sigma_{2}(H_{0})$ respectively. In panel C we observe the appearance of a new regularity region (KAM tori), besides the other tori that we already had in panel A. These invariant curves are organized around a stable periodic orbit family that is born in the system for an energy  $H_{0} = -0.152$, which is surprisingly very close to that of the VRI point. Inter-well transport in this situation is also controlled by the homoclinic intersections that occur between the stable and unstable manifolds of the UPO.

To finish this case we would like to point out that the family of UPOs associated to the  index-1 saddle that sits between both wells undergoes a pitchfork bifurcation just before the energy of the system reaches that of the upper index-1 saddle ($H_0 = 0$), giving rise to two new families of UPOs that we call the top and bottom UPOs because they are located in the well regions of the PES. For more details on the nature of this type of bifurcation of periodic orbits, see the appendix of \cite{katsanikas2018}. This means that for energy values above those of the bifurcation, two new families of UPOs are born in the system, and as we will see in the next subsection, these periodic orbits are responsible for governing the selectivity mechanism in the phase space of the system.

\subsection{Third Case}

In this case, we will focus on describing the system dynamics for an energy range which is above that of the upper index-1 saddle. In particular, we will take a look at the energy value $H_{0}=0.1$. This means that the phase space  bottleneck in the vicinity of the upper index-1 saddle is now open and therefore transport of trajectories between the region of the upper index-1 saddle, the wells and the lower index-1 saddle is allowed. 

\begin{figure}[htbp]
	\begin{center}
		\includegraphics[scale=0.3]{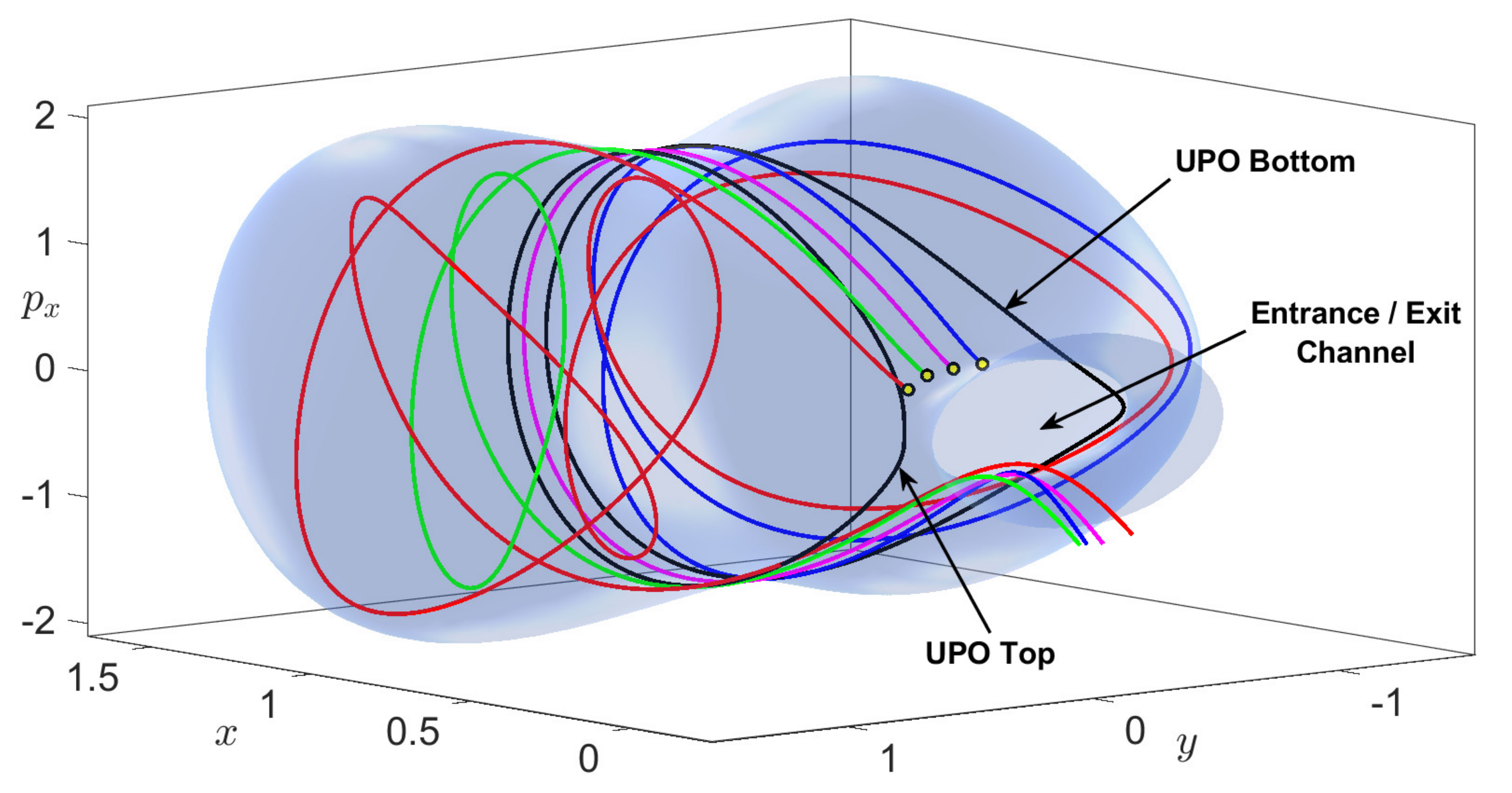}
	\end{center}
	\caption{Dynamical evolution in forward time of four different initial conditions chosen on the surface of section $\Sigma_1(H_0)$, where the system's energy is set to $H_0 = 0.1$. The boundary of the three-dimensional energy hypersurface is shown in blue, and the unstable periodic orbits that control the access of trajectories to the phase space regions corresponding to the potential wells of the PES are depicted in black.}
	\label{trajs_3D}
\end{figure}

Our goal is to describe in detail the mechanisms of transport and trapping of the trajectories that take place in the system, and also the phase space structures responsible for controlling those mechanisms. In Fig. \ref{trajs_3D} we illustrate the energy hypersurface of the system for $H_0 = 0.1$, and we select four different initial conditions (marked with yellow dots) at the entrance channel. We also show the forward evolution of their trajectories in red, green, magenta and blue, and the two black curves representing the unstable periodic orbits that exist in the system in the regions of the wells. In our discussion we will label these UPOs as top and bottom, depending on the well region they are located at. Looking at the behavior of the trajectories in Fig. \ref{trajs_3D} we can easily come to the conclusion that the dynamics of the system is very rich. In this situation, the method of Lagrangian descriptors is very useful, because it provides us with the advantage of easily unveiling the phase space structures that govern the transport mechanism, and allows us to locate the lobes formed by the different stable and unstable manifolds in the system that explain how transport takes place in phase space between different regions of the energy manifold. 

\begin{figure}[htbp]
	\begin{center}
		A)\includegraphics[scale=0.31]{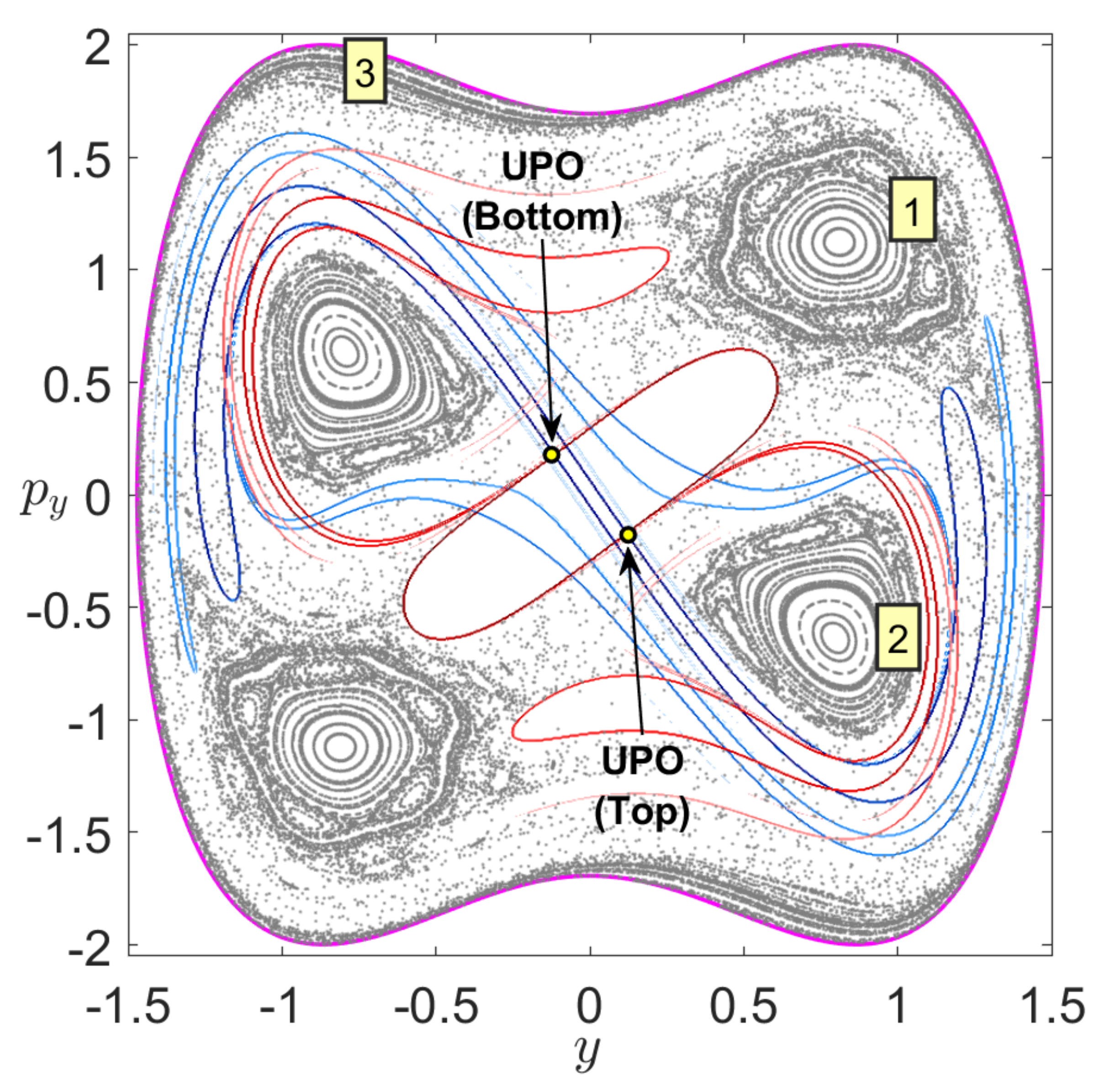}
		B)\includegraphics[scale=0.39]{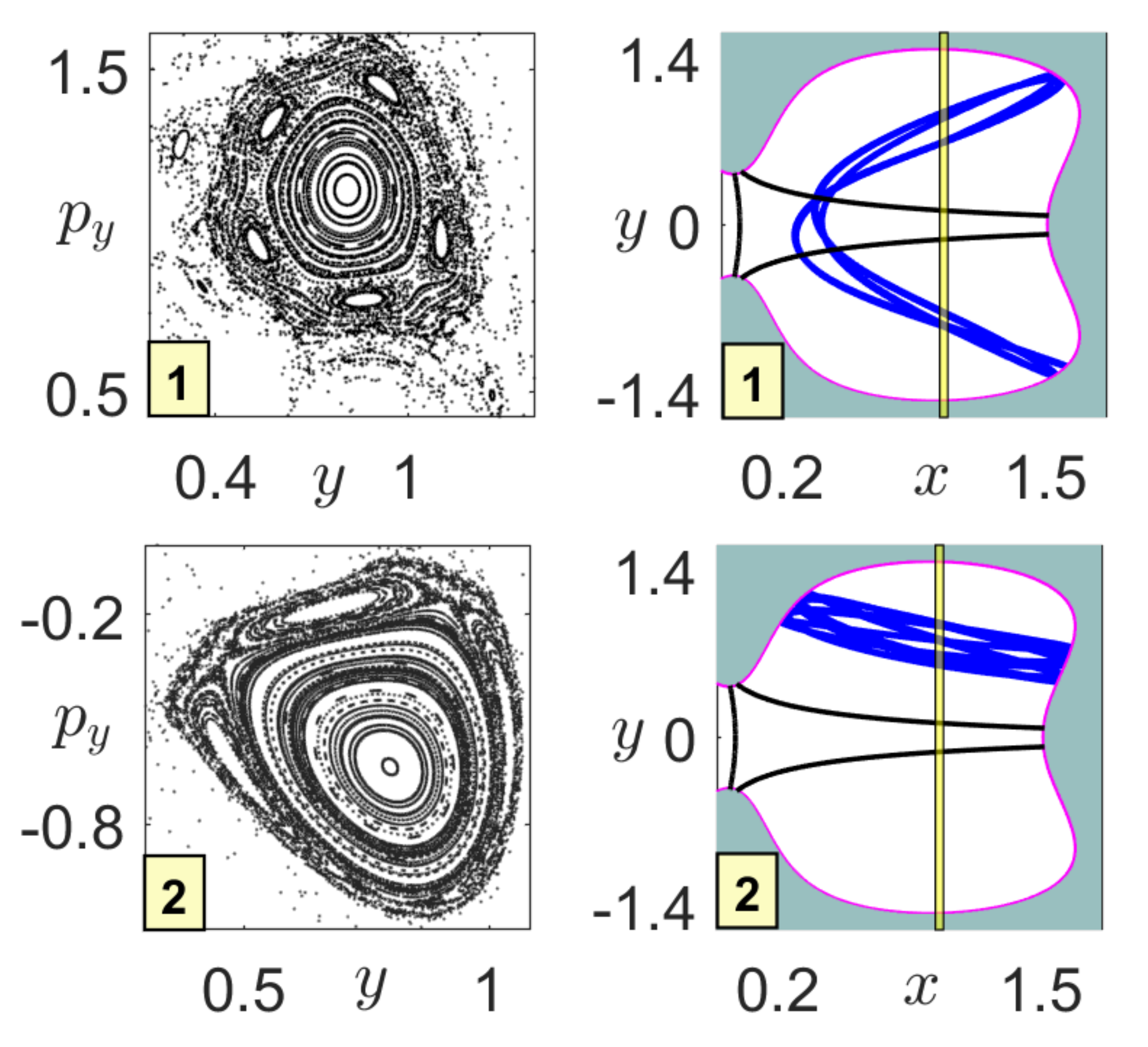}	
		C)\includegraphics[scale=0.3]{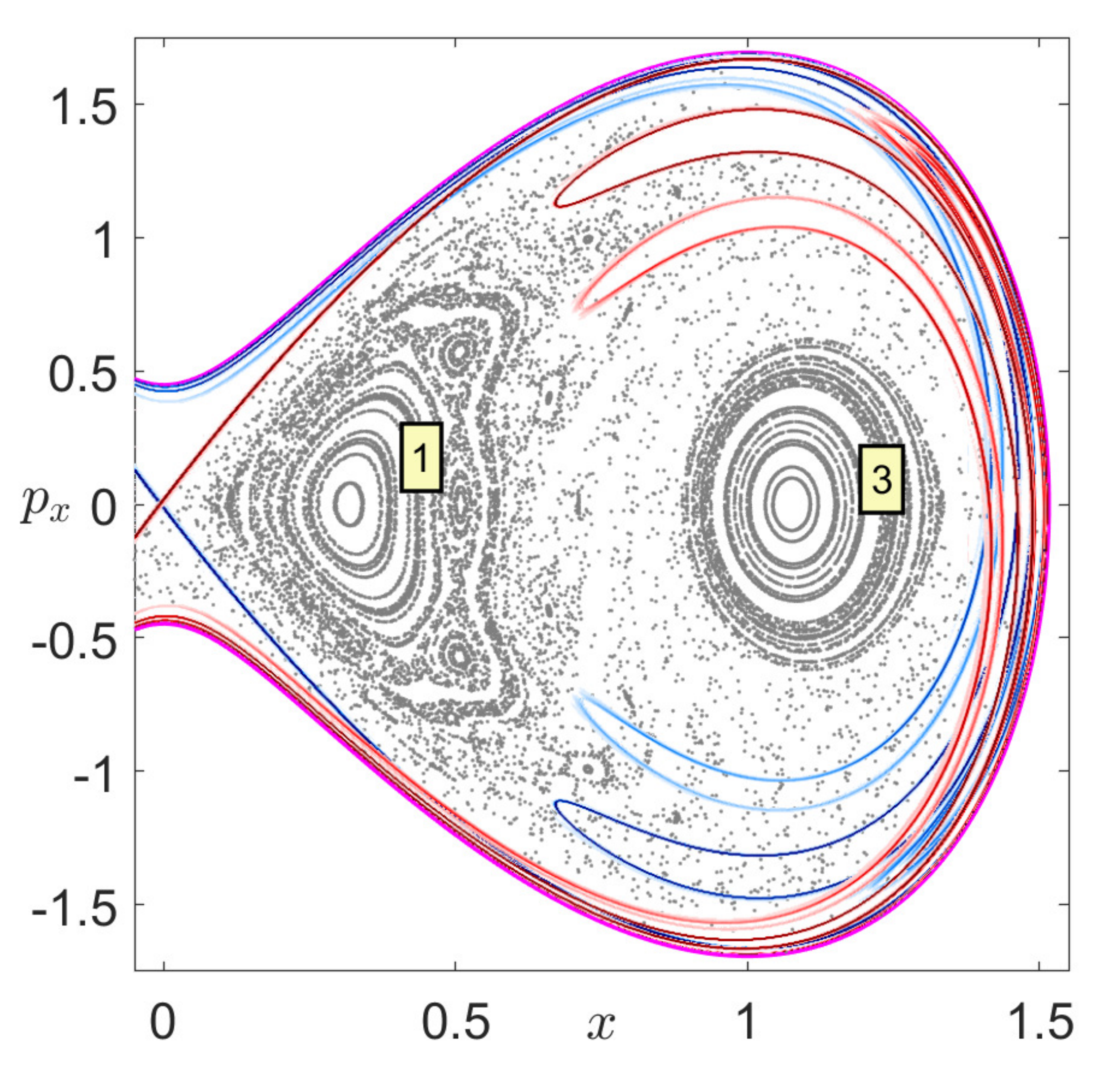}
		D)\includegraphics[scale=0.31]{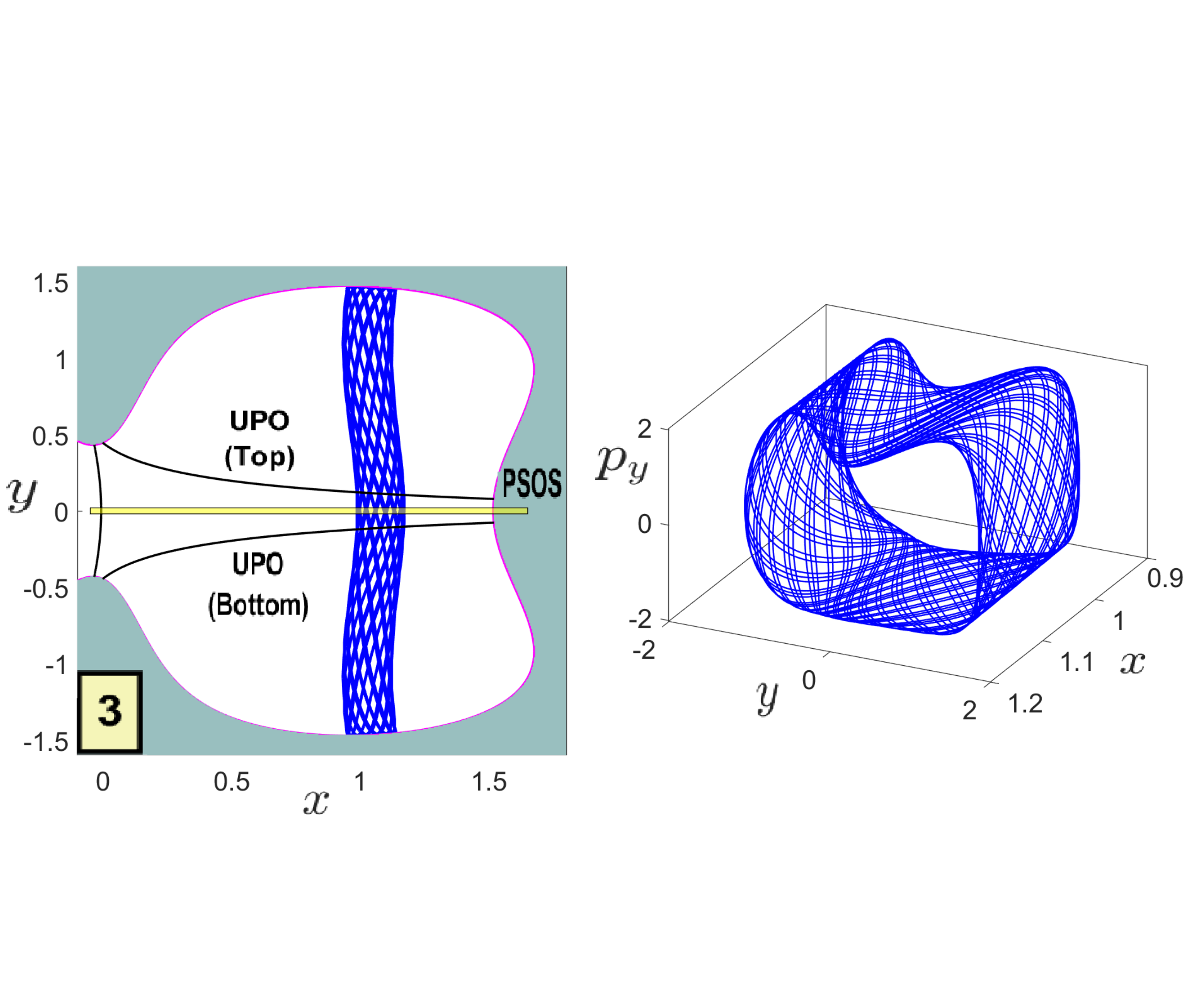}
	\end{center}
	\caption{Regularity behavior displayed by the Hamiltonian system through the KAM tori structures present in its phase space. The energy of the system is chosen as $H_0 = 0.1$. On panels A and C we show Poincar\'e
	maps superimposed with the stable (blue) and unstable (red) manifolds extracted from LDs using $\tau = 8$, and the sections used for the analysis of the dynamics are $\Sigma_2(H_0)$ for A) and $\Sigma_3(H_0)$ for C). On panels B and D, we depict the forward time evolution projected onto configuration space, of three different trajectories corresponding initial conditions labeled on panels A and C. Black curves represent the projections of the UPOs of the system. We also provide a detailed zoom of the regularity regions to illustrate the arrangement of the KAM islands, and for promoting a visual understanding of the tori, we include a three-dimensional representation of one of the torus.}
	\label{en_as_sos_ld}
\end{figure}

We take a look first at the system's dynamics and the phase space structures that characterize transport on the surface of section $\Sigma_{2} (H_{0})$. We observe in Fig. \ref{en_as_sos_ld} A) the invariant curves, labelled as 2, that represent KAM tori around the stable periodic orbits of the families of the top and bottom wells in the central area of the figure. The motion of the trajectories lying on these tori is restricted to a specific region of the PES, which explains why these trajectories are trapped forever in either the top or the bottom well. We illustrate this dynamical behavior in the lower right part of Fig. \ref{en_as_sos_ld} B). Tori that are represented by the invariant curves, labelled by 1, surround the stable periodic orbits of the family of the wells with period 2, and they are inside the lobes formed by the stable and unstable manifolds of the top and bottom UPOs. Trajectories on these tori visit both wells. We display this behavior in the upper right part of Fig. \ref{en_as_sos_ld} B). In Fig. \ref{en_as_sos_ld} C) we see, besides the KAM tori around the stable periodic orbits of the families of the wells (as in panel A), other invariant curves, marked as 3, around the stable periodic orbits of the family that is born in the system close to the energy of the VRI point, Trajectories on these tori visit both wells, see the left part of Fig. \ref{en_as_sos_ld} D).

\begin{figure}[htbp]
	\begin{center}
	    A)\includegraphics[scale=0.28]{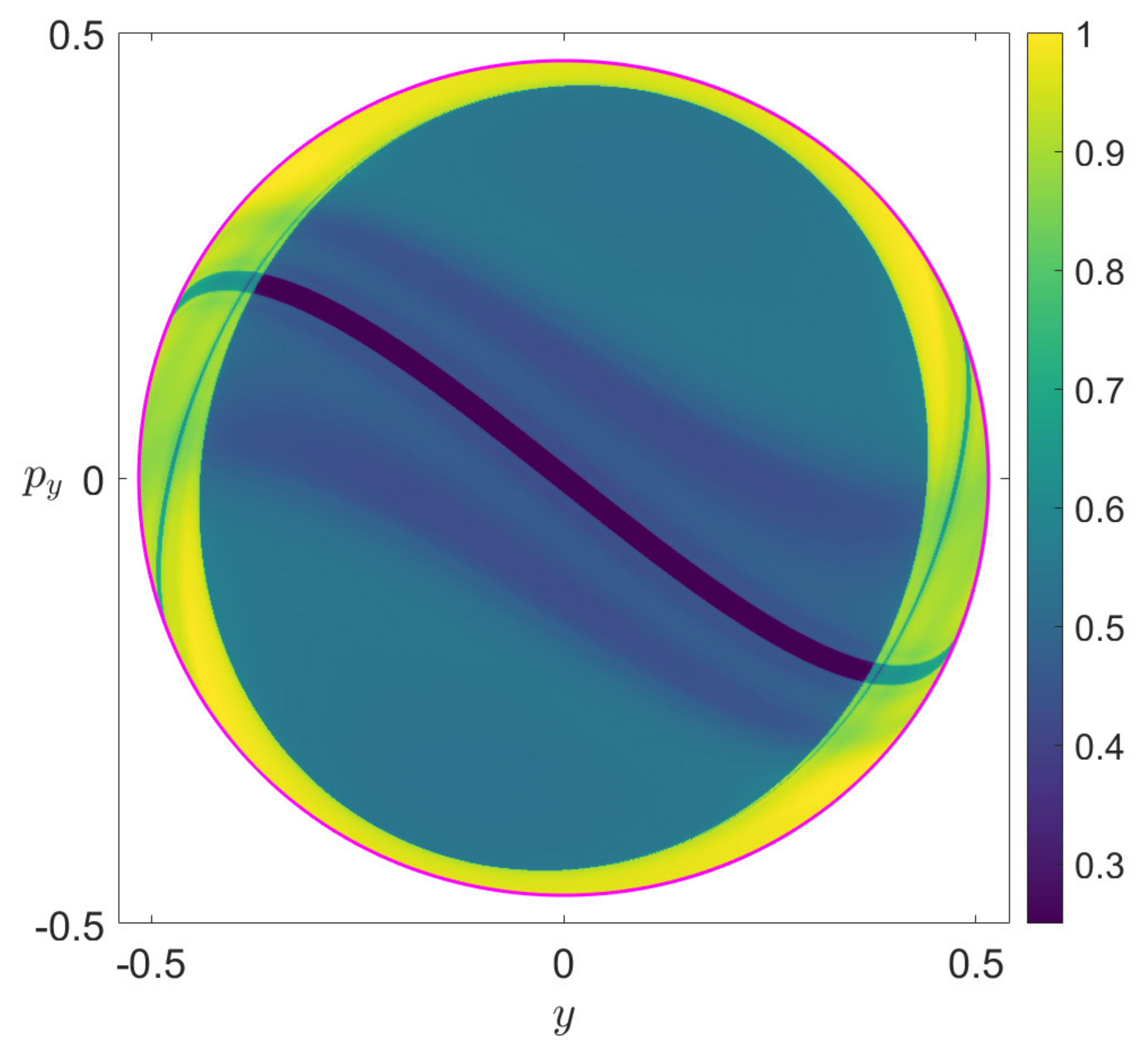}
		B)\includegraphics[scale=0.28]{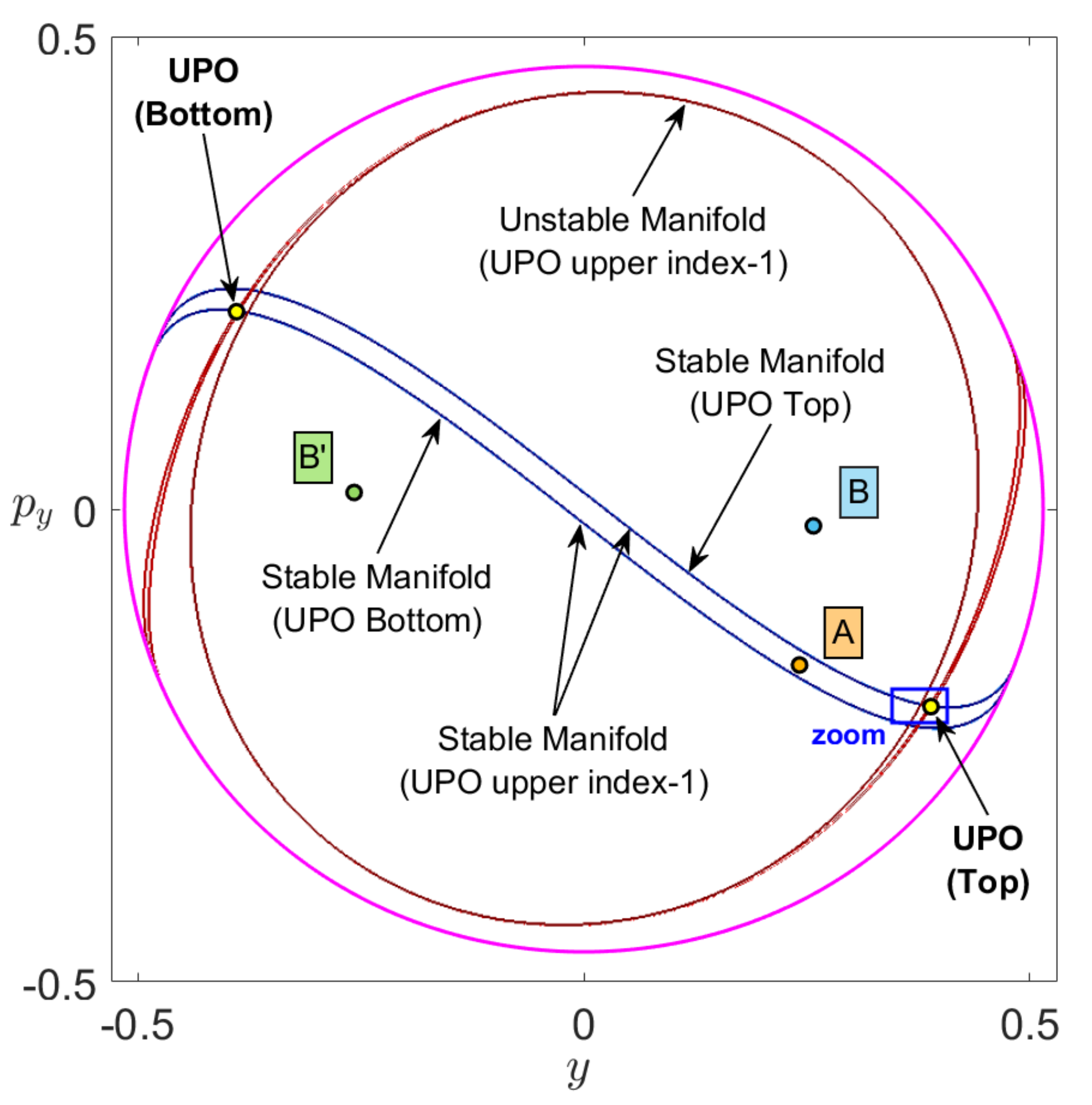}
		C)\includegraphics[scale=0.3]{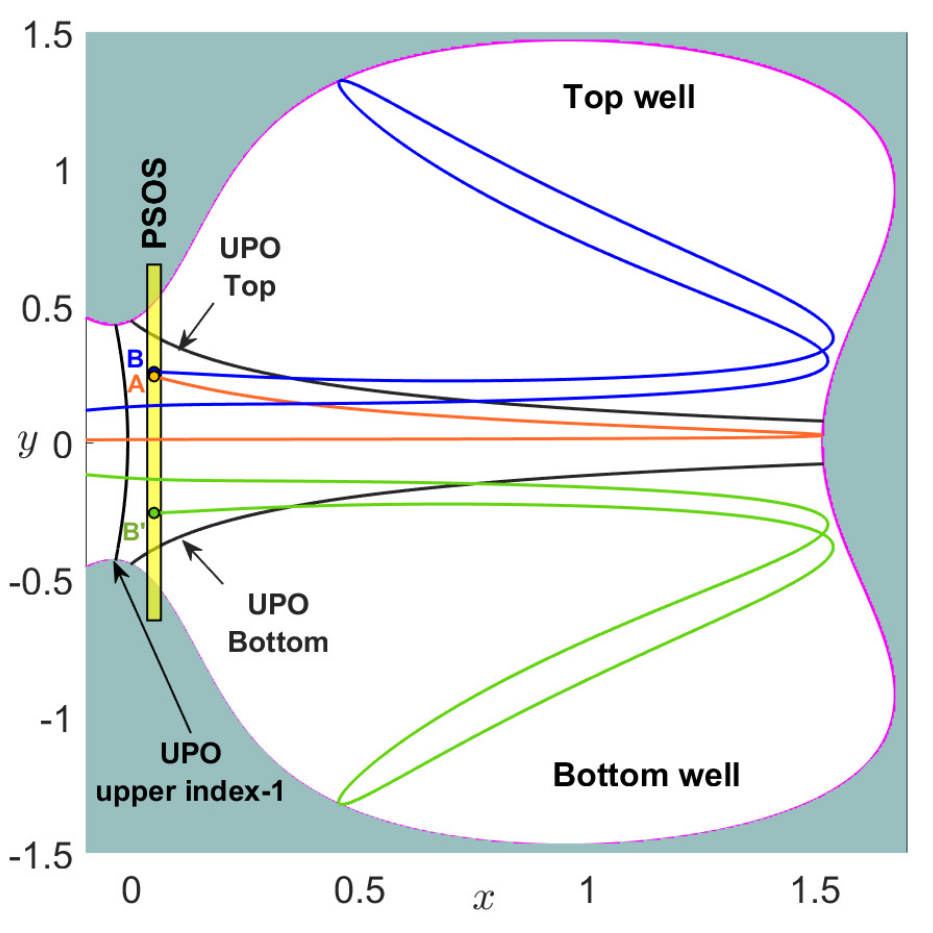}
	\end{center}
	\caption{A) Lagrangian descriptors calculated on the surface of section $\Sigma_1(H_0)$ using $\tau = 6$ for the system's energy $H_0 = 0.1$. B) Stable (blue) and unstable (red) invariant manifolds extracted from the gradient of the LD scalar field. Three different types of initial conditions, labelled as A,B,B', are selected in order to illustrate the symmetry of the lobes in the phase space of the system, and also the basic transport mechanisms governed by the heteroclinic and homoclinic connections. C) Forward time evolution of the trajectories corresponding to the initial conditions chosen in panel B), projected onto configuration space.}
	\label{ld_maniExt}
\end{figure}

\begin{figure}[htbp]
	\begin{center}
		A)\includegraphics[scale=0.36]{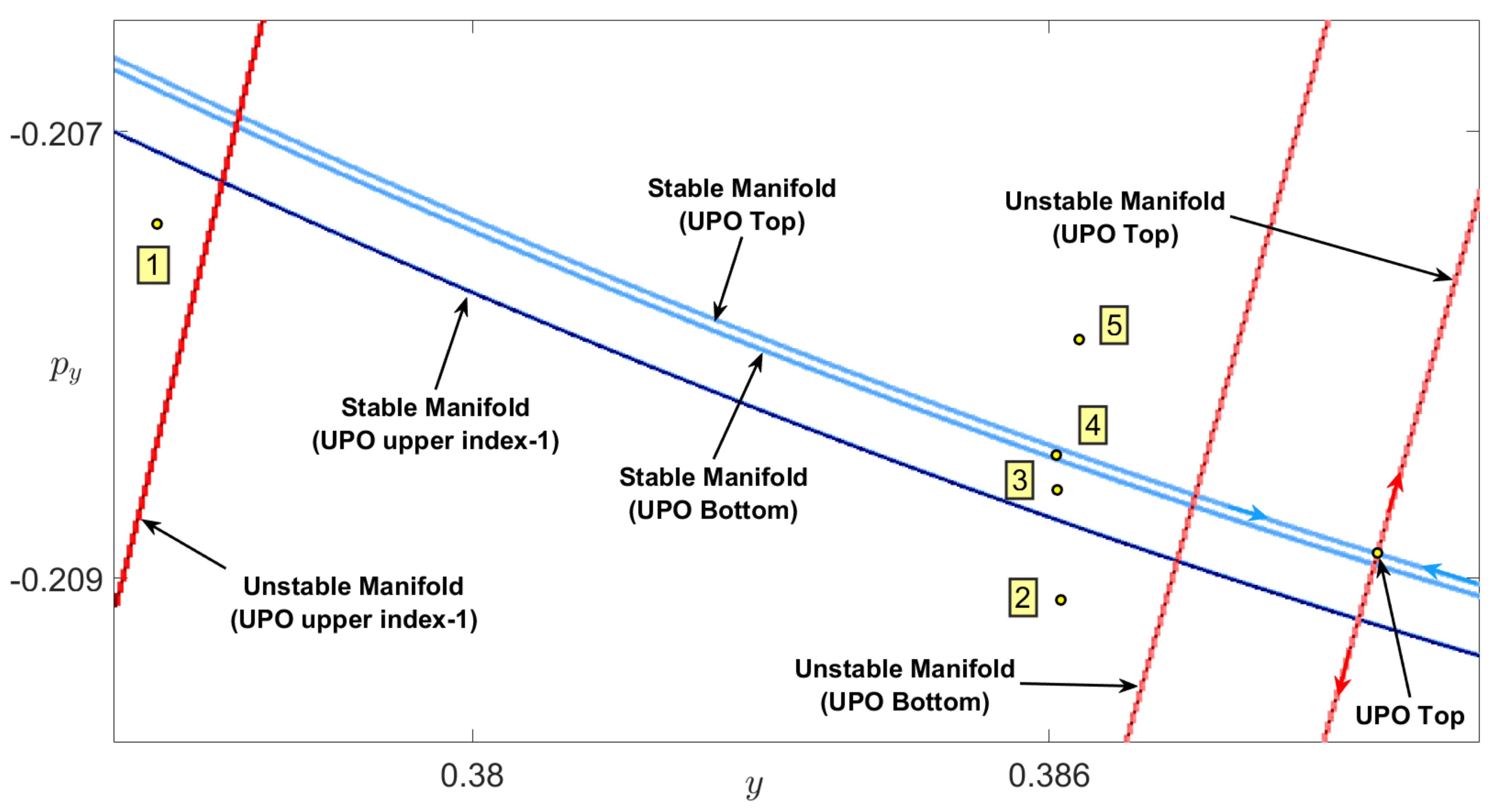} \\
		B)\includegraphics[scale=0.33]{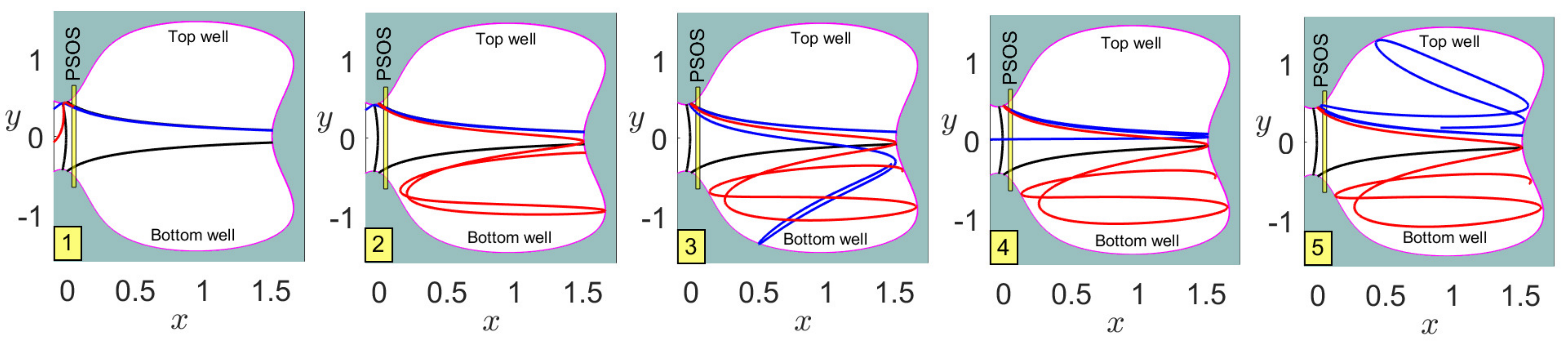}
	\end{center}
	\caption{Lobe dynamics and evolution of trajectories analyzed from the interaction of the stable (blue) and unstable (red) manifolds of the system in a phase space region which corresponds to a zoom of Fig. \ref{ld_maniExt}. A) Invariant manifolds have been extracted from the gradient of LDs calculated on the PSOS $\Sigma_1(H_0)$ using $\tau = 8$ for the system's energy $H_0 = 0.1$. We have selected different initial conditions, marked as yellow dots and labelled from $1$ to $5$ in order to probe the dynamical behavior of lobes. B) Trajectory evolution of the initial conditions in forward (blue) and backward (red) time, projected onto configuration space. The UPOs of the system are also depicted as black curves.}
	\label{lobe_dyn_zoom}
\end{figure}

We move on to investigate the mechanisms of transport of the trajectories in the phase space of the system. To do so, we calculate LDs on the slice $\Sigma_{1}(H_{0})$. In Fig. \ref{ld_maniExt} A) we depict the scalar field obtained from LDs, and in panel B) we show the location of the unstable (red) and stable (blue) manifolds of the different UPOs in the system. We can see that the manifolds of the UPO associated to the upper index-1 saddle, which controls the entrance and exit of trajectories in and out of the PES through the channel, interact with the manifolds of the bottom and top UPOs forming lobes. In order to illustrate the dynamical evolution of trajectories in the regions defined by the lobes, we select three initial conditions that we label by A, B, B$'$. We evolve them forward in time and represent their projections onto configuration space in Fig. \ref{ld_maniExt} C). The initial condition that corresponds to label $A$ is inside a lobe that is associated with the homoclinic intersections of the invariant manifolds of the UPO corresponding to the upper index-1 saddle (see panel B of Fig. \ref{ld_maniExt}). This means that the trajectory follows initially the unstable manifold, getting away from the entrance channel, then it is guided through the homoclinic intersections evolving in the region that lies between both wells until it bounces off the PES wall opposite to the entrance channel and exits the system through the channel without entering any of the well regions. Another example of this type of trajectory behaviour is illustrated by the initial condition labelled as $1$ in Fig. \ref{lobe_dyn_zoom}. On the other hand, the initial conditions B and B$'$ correspond to trajectories located in the lobe that is associated with a heteroclinic intersection between the unstable manifold of the UPO of the upper index-1 saddle with the stable manifold of the top UPO (or the stable manifold of the bottom UPO in the case of B$'$). This means that the trajectories begin from the region of the upper index-1 saddle and they evolve along the unstable manifold of the upper index-1 saddle until they start to follow, through a heteroclinic intersection, the stable manifold of the top or bottom UPOs, entering the region of the top or bottom well respectively. We show this behavior in Fig. \ref{ld_maniExt} C). The symmetric transport of trajectories that we observe is a consequence of the symmetry of the PES with respect to the $y$ coordinate. This property induces a $180^{\circ}$ rotational symmetry about the origin in the phase space that affects the $y$ variable and its canonically conjugate momentum $p_y$. This symmetry is clearly visible in Fig. \ref{ld_maniExt}) B), where the arrangement of the phase space structures and the lobes formed by the stable and unstable manifolds of the UPOs in the system nicely displays this feature. Consequently, the transport mechanism that is responsible for the evolution of trajectories from the entrance channel region of the PES to the region of the top well (or the bottom well) is symmetric, which explains why the branching ratio in this system is $1:1$. What we mean by this is that, from all the trajectories that enter the system through the region of the upper index-1 and visit any of the wells along their evolution, half of them enter first the top well and the other half does the same for the bottom well. We illustrate in Fig. \ref{ld_maniExt}) C) this result and the symmetry property of the phase space structures we discussed above by means of the initial conditions B and B$'$.

We turn our attention next to secondary transport mechanisms that take place in the phase space of the system. We illustrate them by selecting initial conditions in different regions of the phase space characterized by the lobes formed by the interactions of the stable and unstable manifolds of the UPOs. In particular, we focus on the initial conditions labelled $2$, $3$ and $5$ in Fig. \ref{lobe_dyn_zoom} A), which correspond to three qualitatively distinct mechanisms of transport:
\begin{enumerate}
    \item \underline{{\bf First mechanism:}} This mechanism is responsible for the transport of trajectories from the region of the bottom or top wells to the region of the exit channel. In this mechanism, the trajectories that are located in the region of the top or bottom well, follow the unstable manifolds of the top or bottom UPOs which have heteroclinic intersections with the stable manifold of the UPO associated to the upper index-1 saddle. Then, the trajectories are guided through these heteroclinic intersection and evolve from the region of the wells to the region of the exit channel of the PES. The trajectory that starts from the initial condition 2 in Fig. \ref{lobe_dyn_zoom} A) is a representative example of this mechanism. This initial condition is located inside a lobe that is associated with the heteroclinic intersection between the unstable manifold of the bottom UPO with the stable manifold of the UPO of the upper index-1 saddle. The trajectory is coming from the region of the bottom well, and it moves towards the exit channel where it escapes the system. This dynamical behavior is depicted in the second panel Fig. \ref{lobe_dyn_zoom}) B).
     
     \item \underline{{\bf Second mechanism:}} This mechanism is characterized by the homoclinic intersections between the stable and unstable manifolds of the top UPO (or the bottom UPO). We explain it for the bottom UPO, because the mechanism is exactly the same for the top UPO due to the symmetry in the system. In this situation, trajectories follow initially the unstable manifold of the bottom UPO, moving away of the bottom well region, and through homoclinic intersections with the the stable manifold of the bottom UPO, they come back to the bottom well. An example of a trajectory that displays this behaviour is given by the initial condition 3 in Fig.\ref{lobe_dyn_zoom} A). This trajectory is located inside a lobe that is associated with the homoclinic intersection of the invariant manifolds of the bottom UPO. The trajectory starts from the region of the bottom well, moves to the region between both wells without visiting the top well, and then it returns to the region of the bottom well. This is depicted in the third panel of Fig. \ref{lobe_dyn_zoom} B). 
    
     \item \underline{{\bf Third mechanism:}} This mechanism is responsible for the transport of trajectories between the regions of the two wells (inter-well transport). In this situation, trajectories that are located initially in the region of one of the two wells, for example the top well, follow the unstable manifolds of the top UPO, and through heteroclinic intersections with the stable manifolds of the bottom UPO, visit the bottom well. An example of such behavior given by the trajectory starting from the initial condition 5, see Fig. \ref{lobe_dyn_zoom} A). This trajectory has an initial condition inside a lobe associated with the heteroclinic intersection of the unstable manifold of the bottom UPO with the stable manifold of the top UPO. The trajectory starts from the region of the bottom well, and it evolves in a way that it enters the top well, see the last panel Fig. \ref{lobe_dyn_zoom} B).
\end{enumerate}

\begin{figure}[htbp]
	\begin{center}
		A)\includegraphics[scale=0.3]{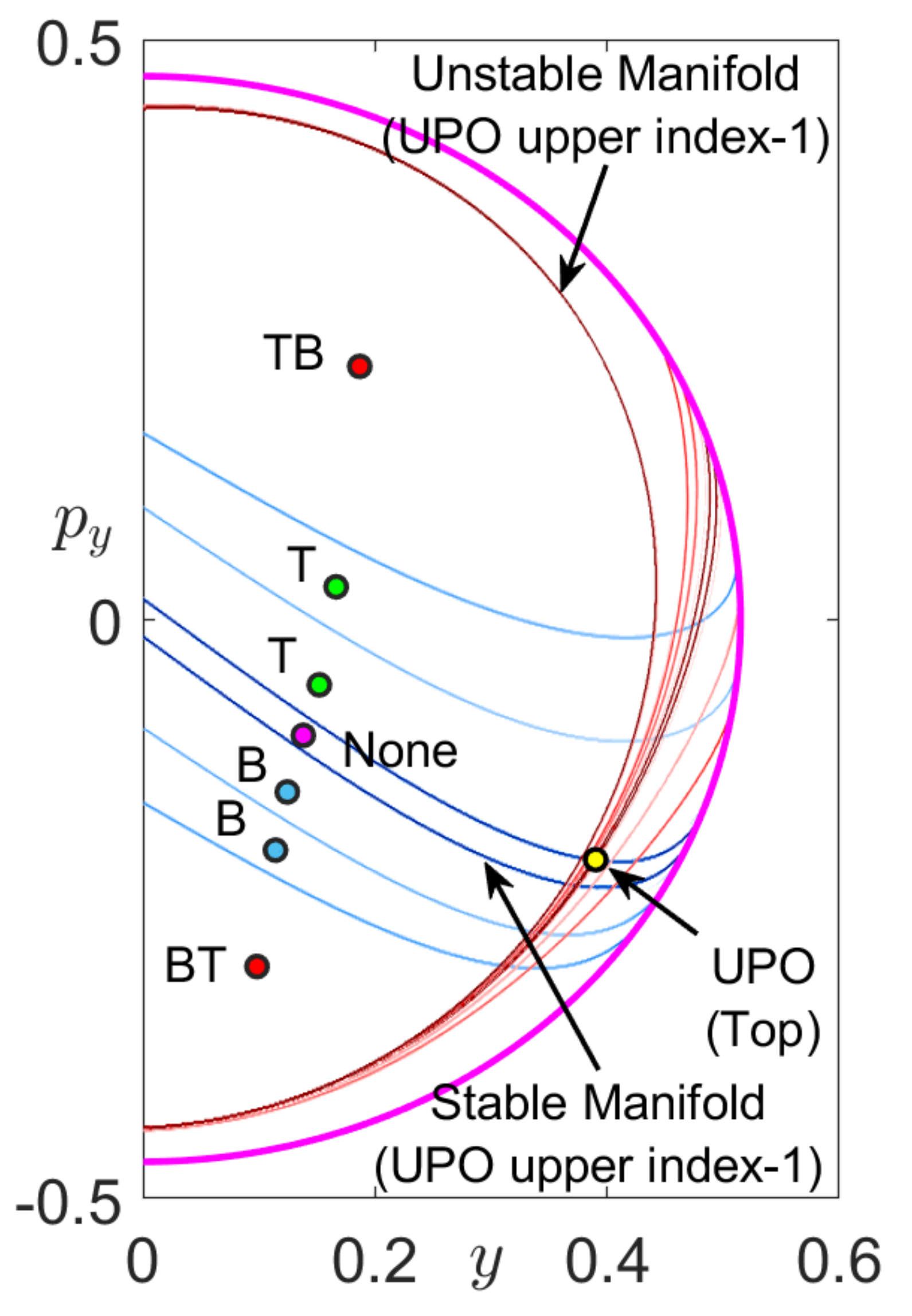}
		B)\includegraphics[scale=0.3]{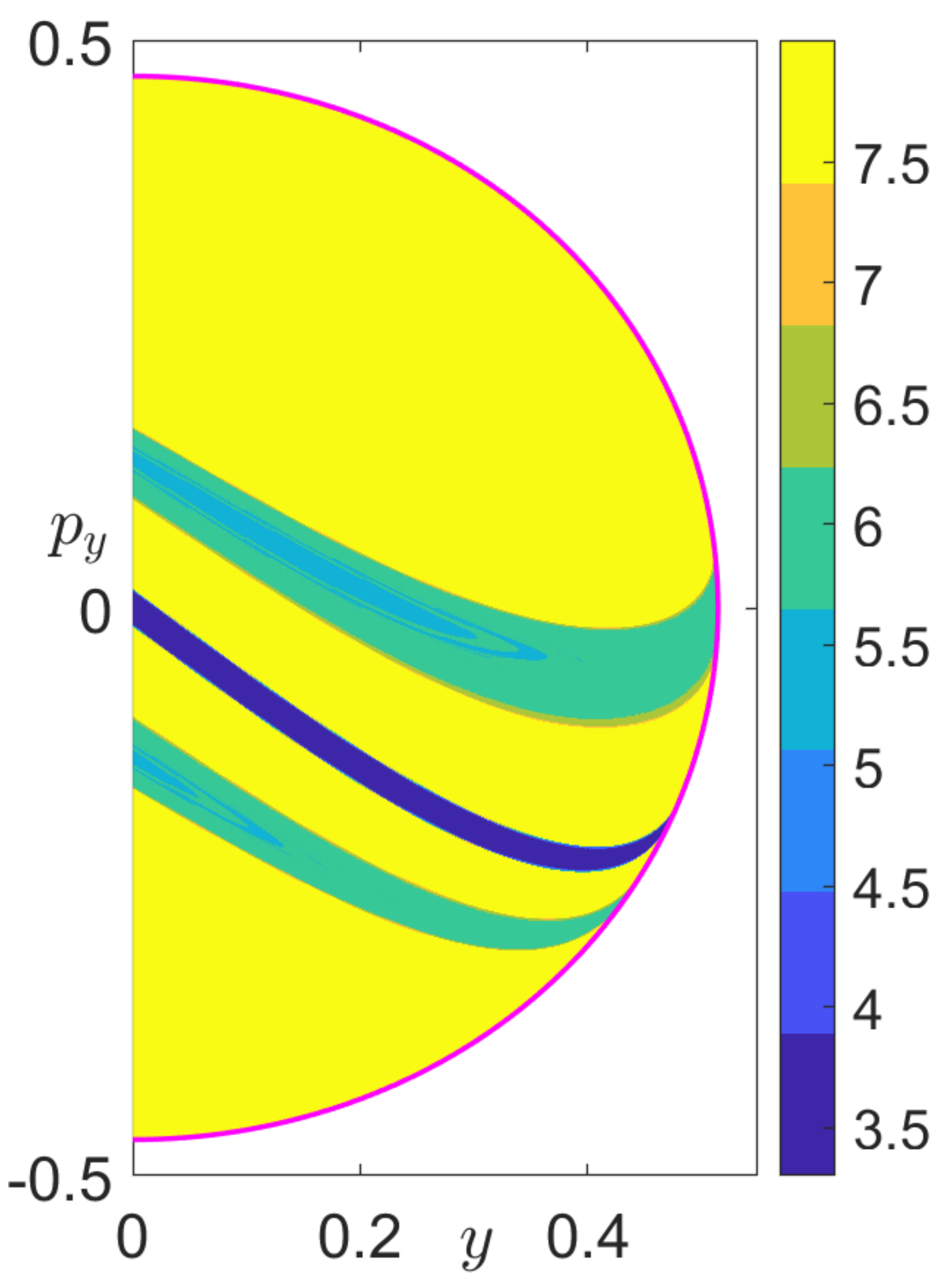}
		C)\includegraphics[scale=0.3]{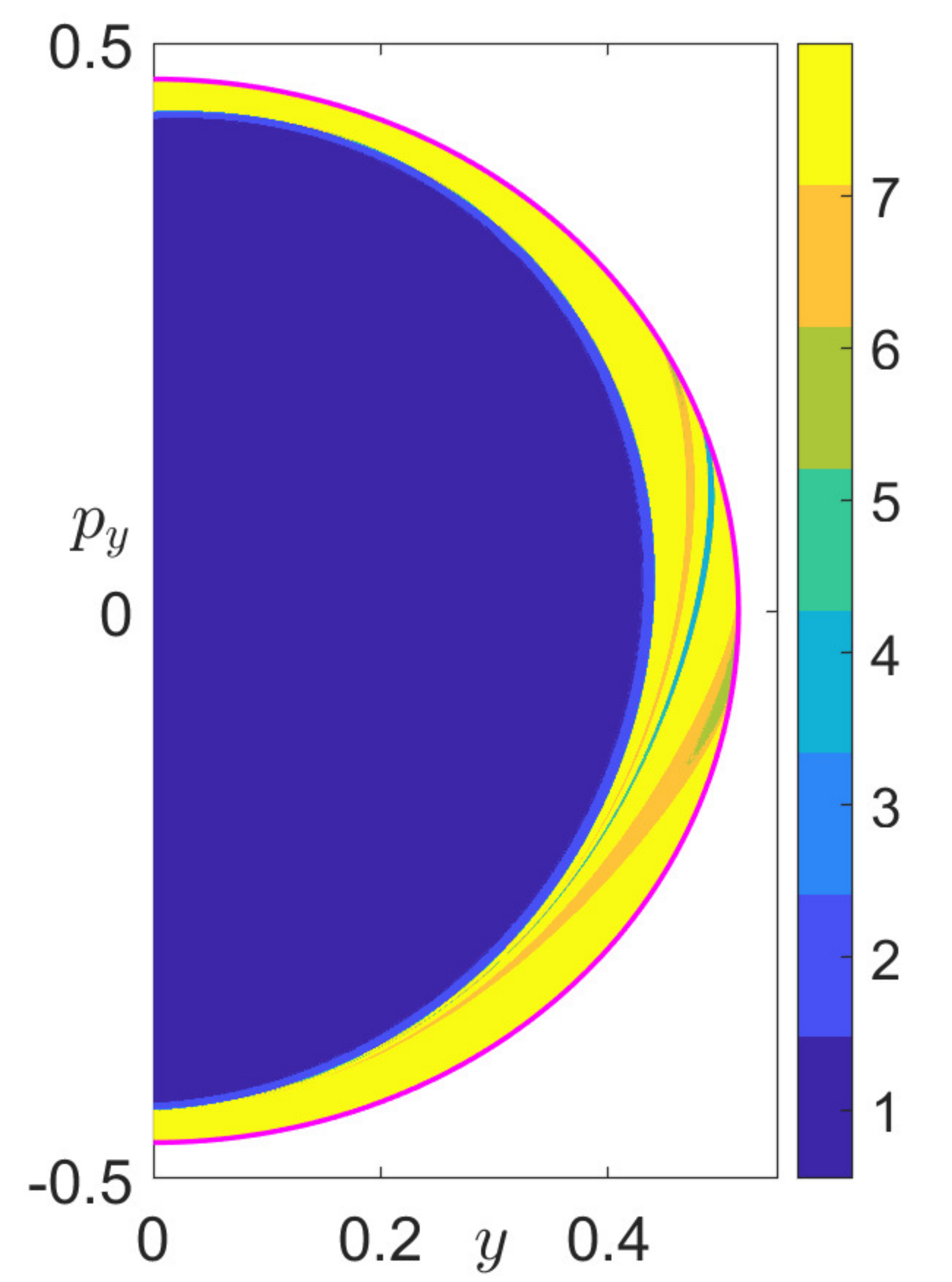}
		D)\includegraphics[scale=0.42]{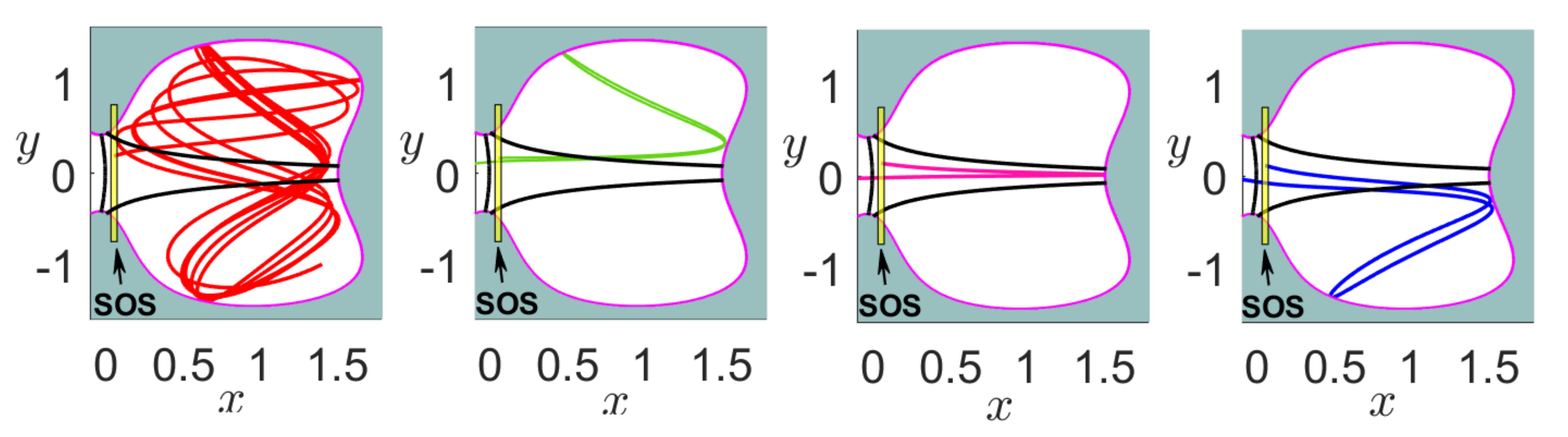}
	\end{center}
	\caption{Analysis of lobe dynamics, escape times and forward time evolution of trajectories for the system with energy $H_0 = 0.1$ on the surface of section $\Sigma_1(H_0)$. A) Stable (blue) and unstable (red) manifolds extracted from the LD scalar field calculated using $\tau = 8$. We have marked initial conditions with circles of different colors in order to probe the system's dynamics. The evolution of their trajectories, using the same color scheme, is shown in panel D). An explanation of the labels (T)-(B)-(TB)-(BT)-(None) is given in the main text of the paper; B) Forward escape time plot; C) Backward escape time plot; D) Trajectory evolution of the initial conditions selected in panel A), projected onto configuration space. We have also depicted with black curves the projections of the UPOs of the system.}
	\label{lobe_dyn} 
\end{figure}

At this point, it is important to highlight that due to the symmetry of the PES with respect to the $y$ coordinate, phase space transport displays a $180^{\circ}$ rotational symmetry about the origin that involves both $y$ and its conjugate momentum $p_y$. This property determines the dynamical fate of trajectories in secondary lobes that are related by this symmetry in the system. This means that if we choose one initial condition in a lobe that visits the top well, the symmetric initial condition will visit the bottom well. In order to confirm the symmetric dynamical behavior of our system we have chosen seven different initial conditions in the surface of section $\Sigma _{1}(H_{0})$, see Fig. \ref{lobe_dyn} A). We have integrated these trajectories in forward time, see panel D for their projections onto configuration space, and labelled them according to their dynamical behavior. The label TB stands for trajectories that move from the top well to the bottom well, BT stands for trajectories that move from the bottom well to the top well, T indicates trajectories that visit only the top well, B mark those trajectories that visit only the bottom well, and None stands for the trajectories that do not visit any of the wells. Notice that due to the symmetry in the PES we only need to focus on positive values of the $y$ coordinate, which corresponds to the top well region. It is important to remark here that if we choose an initial condition in a lobe that visits the top well the symmetric initial condition will be in a lobe associated to trajectories that visit the bottom well. An example of this behavior is provided by the red initial conditions labelled TB and BT. In this case, the trajectory with the red initial condition TB visits the top and then the  bottom well. If we choose the symmetric initial condition, marked as BT, and constructed by doing a $180^{\circ}$ rotational symmetry about the origin, this initial condition will be located in a lobe in which all trajectories will visit the bottom and then the top well. Other examples of the effect that the symmetry has on the transport mechanism in the system is displayed by the the green and blue initial conditions. The trajectory followed by the green initial condition visits the top well and then it exits the system. On the other hand, the trajectory of the blue initial condition visits the bottom well and then it escapes to infinity through the entrance channel of the PES. For the magenta initial condition, we have an example of a trajectory that does not visit any of the two wells. This type of trajectory bounces off the wall of the PES that is opposite to the entrance channel and exits the system.

We want to underline the fact that the size of the lobes is very important for the transport. An example of the influence of the size of lobes is provided by the initial condition 4, see Fig. \ref{lobe_dyn_zoom} A). This initial condition is inside a lobe that is associated with a heteroclinic intersection of the unstable  manifold of the bottom UPO and stable manifold of the top UPO.
Moreover, it is also located in a large lobe formed by the unstable manifold of the bottom UPO and the stable manifold of the UPO corresponding to the upper index-1 saddle. Therefore, the trajectory will evolve initially from the bottom well region towards the top well, but  it will not have enough time to enter the region of the top well, since the larger lobe will dominate the lobe dynamics, making the trajectory escape through the exit channel. On the other hand, the trajectory associated to the initial condition 5 is located in a large lobe, see Fig. \ref{lobe_dyn_zoom}. We discussed this case above as an example of the third type of secondary mechanism. 

Now we will discuss the escape time of the trajectories and the influence of the size of the lobes on this property. In panel B of Fig. \ref{lobe_dyn} we illustrate the exit times forwards in time. It is evident that the trajectories that escape faster belong to the lobes that do not interact with any of the wells. Moreover the trajectories that escape slower belong to the lobes that visit both wells. This happens because these lobes are larger than the others and the trajectories are trapped for longer times inside them. In panel C we present the exit times backwards in time, where it is clear that the trajectories that escape faster are those that are inside the unstable manifold of the UPO of the upper index-1 saddle.

\section{Conclusions}
\label{sec:conc}

In this work we have studied, by means of combining the method of Lagrangian descriptors with the classical approach of Poincar\'e maps, the phase space dynamics of a Hamiltonian system with 2 DoF defined by a symmetric PES with an entrance/exit channel determined by a high energy index-1 saddle and two potential wells separated by a low energy saddle. This benchmark model has provided us with a testbed to explain how selectivity arises naturally as a dynamical mechanism in the phase space of the system. This is important because it allows us to develop a fundamental understanding of this phenomenon, which is relevant for the analysis of product distributions in chemical reaction dynamics.

Our analysis has revealed that the branching of trajectories that enter the system through the phase space bottleneck of the entrance/exit channel is controlled by the heteroclinic intersections established between the unstable manifold of the UPO associated to the upper index-1 saddle, and the stable manifolds of the two families of UPOs that exist in the regions of the wells. These heteroclinic connections are responsible for guiding the trajectories towards either well in the system. Moreover, by means of a stability analysis we have found that the top and bottom UPOs in the regions of the wells are generated through a pitchfork bifurcation that occurs in the family of UPOs associated to the lower index-1 saddle, and this happens for an energy level just below that of the upper index-1 saddle. This means that when the phase space bottleneck of the entrance channel opens, these two families of UPOs are born, and the heteroclinic interactions of their stable manifolds with the unstable manifolds of the UPO corresponding to the upper index-1 saddle govern the branching mechanism in the system. 

Interestingly, in this setup the expected branching ratio would be $1:1$ due to two factors: the symmetry of the PES with respect to the $y$ coordinate, and the symmetric locations of the wells, having both of them the same energy. The computations we have carried out show that the symmetry in the PES induces a rotational symmetry of $180^{\circ}$ about the origin in the phase space that involves the $y$ coordinate and its canonically conjugate momentum $p_y$. This property makes the phase space structures present in the top and bottom well regions of the PES symmetric, and hence this condition enforces the equal branching obtained, since the interactions through lobes occur in exactly the same way.

In addition to studying the selectivity mechanism, we have also investigated well to well transport in the system. For this case, the dynamical mechanism that governs transport is the heteroclinic connections that exist between the unstable invariant manifolds of the top/bottom UPO with the stable invariant manifolds of the bottom/top UPO. Finally, another important transport mechanism that we have explored in this work is that followed by trajectories that enter the system through the entrance channel, bounce off the PES wall opposite to the channel, and escape the system without interacting with either well. The trajectories that display this dynamical behavior are located inside the lobes formed by the homoclinic connections between the unstable and stable invariant manifolds of the unstable periodic orbit associated to the upper index-1 saddle.    

\section*{Acknowledgments}

The authors would like to acknowledge the financial support provided by the EPSRC Grant No. EP/P021123/1 and the Office of Naval Research Grant No. N00014-01-1-0769.

\bibliography{SNreac}

\appendix

%%%%%%%%%%%%%%%%%%%%%%%%%%%%%%%%%%%%%%%%%%%%%%%%%%%%%%%
%%%%%%%%%%%%%%%%%%%  APPENDICES %%%%%%%%%%%%%%%%%%%%%%%
%%%%%%%%%%%%%%%%%%%%%%%%%%%%%%%%%%%%%%%%%%%%%%%%%%%%%%%

\section{Lagrangian Descriptors}
\label{sec:appA}

Consider a dynamical system with general time dependence:
\begin{equation}
\dfrac{d\mathbf{x}}{dt} = \mathbf{v}(\mathbf{x},t) \;,\quad \mathbf{x} \in \mathbb{R}^{n} \;,\; t \in \mathbb{R} \;,
\label{eq:gtp_dynSys}
\end{equation}
where the vector field satisfies $\mathbf{v}(\mathbf{x},t) \in C^{r} (r \geq 1)$ in $\mathbf{x}$ and continuous in time. The natural way to explore phase space structure is to use trajectories, since these objects are its building blocks and the geometry of the underlying phase space is encoded in the initial conditions themselves. The simple and elegant idea behind LDs in order to provide a qualitative description of the system's dynamics is to seed a given phase space region with initial conditions and integrate a bounded and positive quantity (an intrinsic geometrical and/or physical property of the dynamical system under study) along trajectories for a finite time. The definition of LDs that we use in this work follows the one presented in \cite{lopesino2017}, which relies on integrating along trajectories the $p$-norm of the vector field of the dynamical system, where $p \in (0,1]$ is a parameter chosen in advance. In this work, we will use for the simulations the value $p = 1/2$. In the works \cite{demian2017,naik2019a}, a rigorous theoretical foundation for this methodology is established, and a mathematical connection is found between normally hyperbolic invariant manifolds (NHIMs) and their stable and unstable manifolds, and the ``singular structures'' that appear in the LD scalar field.

Given a fixed integration time $\tau > 0$ and let $\mathbf{x}_0 = \mathbf{x}(t_0)$ be any initial condition of the system. We define the fixed-time integration LDs diagnostic calculated at time $t_0$ as:
\begin{equation}
M_p(\mathbf{x}_{0},t_0,\tau) = \sum_{k=1}^{n} \bigg[ \int^{t_0+\tau}_{t_0-\tau}  |v_{k}(\mathbf{x}(t;\mathbf{x}_0),t)|^p \; dt \bigg]  \;, 
\label{eq:Mp_function}
\end{equation}
where $v_k$ is the $k$-th component of the vector field that defines the dynamical system in Eq. \eqref{eq:gtp_dynSys}. Notice that this definition can be decomposed into its forward and backward integration parts:
\begin{equation}
\begin{split}
M_p^{(b)}(\mathbf{x}_{0},t_0,\tau) &= \sum_{k=1}^{n} \bigg[ \int^{t_0}_{t_0-\tau}  |v_{k}(\mathbf{x}(t;\mathbf{x}_0),t)|^p \; dt \bigg] \;, \\[.2cm]
M_p^{(f)}(\mathbf{x}_{0},t_0,\tau) &= \sum_{k=1}^{n} \bigg[ \int^{t_0+\tau}_{t_0} |v_{k}(\mathbf{x}(t;\mathbf{x}_0),t)|^p \; dt \bigg] \;,
\end{split}
\end{equation}
The advantage of splitting function $M_p$ into its forward and backward components is that forward integration highlights the stable manifolds of the dynamical system, while backward evolution recovers the unstable manifolds. Moreover, the combination of both forward and backward detects all the invariant manifolds simultaneously. This detection of invariant manifolds by means of locations at which the LD scalar field becomes non-differentiable has been mathematically quantified in terms of the notion of ``singular structures'' in the LDs plots, which are easy to recognize visually \cite{mancho2013lagrangian,lopesino2017,demian2017,naik2019a}. Therefore, this approach allows us to easily extract the manifolds from the high values (ridges) attained by the gradient of the scalar function itself.

The methodology offered by LDs has thus the capability of producing a complete and detailed geometrical \textit{phase space tomography} in high dimensions by means of using low-dimensional phase space probes to extract the intersections of the phase space invariant manifolds with these slices \cite{demian2017,naik2019a,naik2019b}. Any phase space slice can be selected and sampled with a high-resolution grid of initial conditions, and no information regarding the dynamical skeleton of invariant manifolds at the given slice is lost as the trajectories evolve in time. Moreover, this analysis does not rely on trajectories coming back to the chosen slice, as is required for Poincar\'e maps to work. In this respect, there is also another key point that needs to be highlighted which demonstrates the real potential of LDs with respect to other classical nonlinear dynamics techniques. Using LDs one can obtain \textit{all} the invariant manifolds of the dynamical system \textit{simultaneously}, and this comes with a tremendous save in the computational cost, since LDs are extremely simple and straightforward to implement.

In order to apply LDs for revealing the invariant manifolds in phase space, it is very important to remark the crucial role played by the integration time $\tau$ in the definition of the method. The consequence of increasing the value for $\tau$ is that richer and more intricate details of the underlying geometrical template of phase space structures are unveiled. This is the expected behavior, since an increase of the integration time would imply incorporating more information about the past and future dynamical history of trajectories in the computation of LDs. This means that $\tau$ is  intimately related to the time scales of the dynamical phenomena that occur in the model under consideration. This connection makes the integration time a problem-dependent parameter, and hence, there is no general ``golden rule'' for selecting its value for exploring phase space. One needs to bare in mind in this context that there exists a compromise between the complexity of the structures one would like to reveal from the application of the method in order to explain a certain dynamical mechanism, and the interpretation of the intricate manifolds displayed in the LD scalar output after the simulation is carried out. We will present an example regarding the importance of this property in the next section. 

Since the Hamiltonian system we are dealing with in this work has an unbounded phase space, we need to be careful when applying LDs to reveal its invariant manifolds as trajectories can escape to infinity at a very fast rate or even in finite time. This issue is related to the fact that all initial conditions in the definition of LDs in Eq. \eqref{eq:Mp_function} are integrated for the same time $\tau$. Recent studies have revealed \cite{junginger2017chemical,naik2019b,GG2020a,katsanikas2020b} that this type of trajectory behavior can obscure the detection of invariant manifolds. In order to circumvent this problem, we adapt Eq. \eqref{eq:Mp_function} by adopting here the approach known as variable integration time Lagrangian descriptors. In this methodology, LDs are calculated, at any initial condition, for a fixed initial integration time $\tau_0$ or until the trajectory of that initial condition leaves a certain phase space region $\mathcal{R}$ that we call the {\em interaction region}. Therefore, the total integration time in this strategy depends on the initial conditions themselves, that is $\tau(\mathbf{x}_0)$. In this variable-time formulation, given a fixed integration time $\tau_0 > 0$, the $p$-norm definition of LDs with $p \in (0,1]$ has the form:
\begin{equation}
M_p(\mathbf{x}_{0},t_0,\tau) = \sum_{k=1}^{n} \left[ \int^{t_0 + \tau^{+}_{\mathbf{x}_0}}_{t_0 - \tau^{-}_{\mathbf{x}_0}}  |v_{k}(\mathbf{x}(t;\mathbf{x}_0),t)|^p \; dt \right]  \;,
\label{eq:Mp_vt}
\end{equation}
and the total integration time is defined as:
\begin{equation}
\tau^{\pm}_{\mathbf{x}_{0}} = \min \left\lbrace \tau_0 \, , \, |t^{\pm}|_{\big| \mathbf{x}\left(t^{\pm}; \, \mathbf{x}_{0}\right) \notin \mathcal{R}} \right\rbrace \; ,
\end{equation}
where $t^{+}$ and $t^{-}$ are the times for which the trajectory leaves the interaction region $\mathcal{R}$ in forward and backward time, respectively. For this work we will define the interaction region as:
\begin{equation}
\mathcal{R} = \left\{ \left(x,y,p_x,p_y\right) \in \mathbb{R}^4 \;\Big| \; x > -0.1 \right\} 
\label{eq:intReg}
\end{equation}
which reflects the physical assumption that trajectories of the system escaping through the entrance/exit channel of the PES, characterized by the index-1 saddle at the origin, will never return.

\begin{figure}[htbp]
	\begin{center}
		A)\includegraphics[scale=0.24]{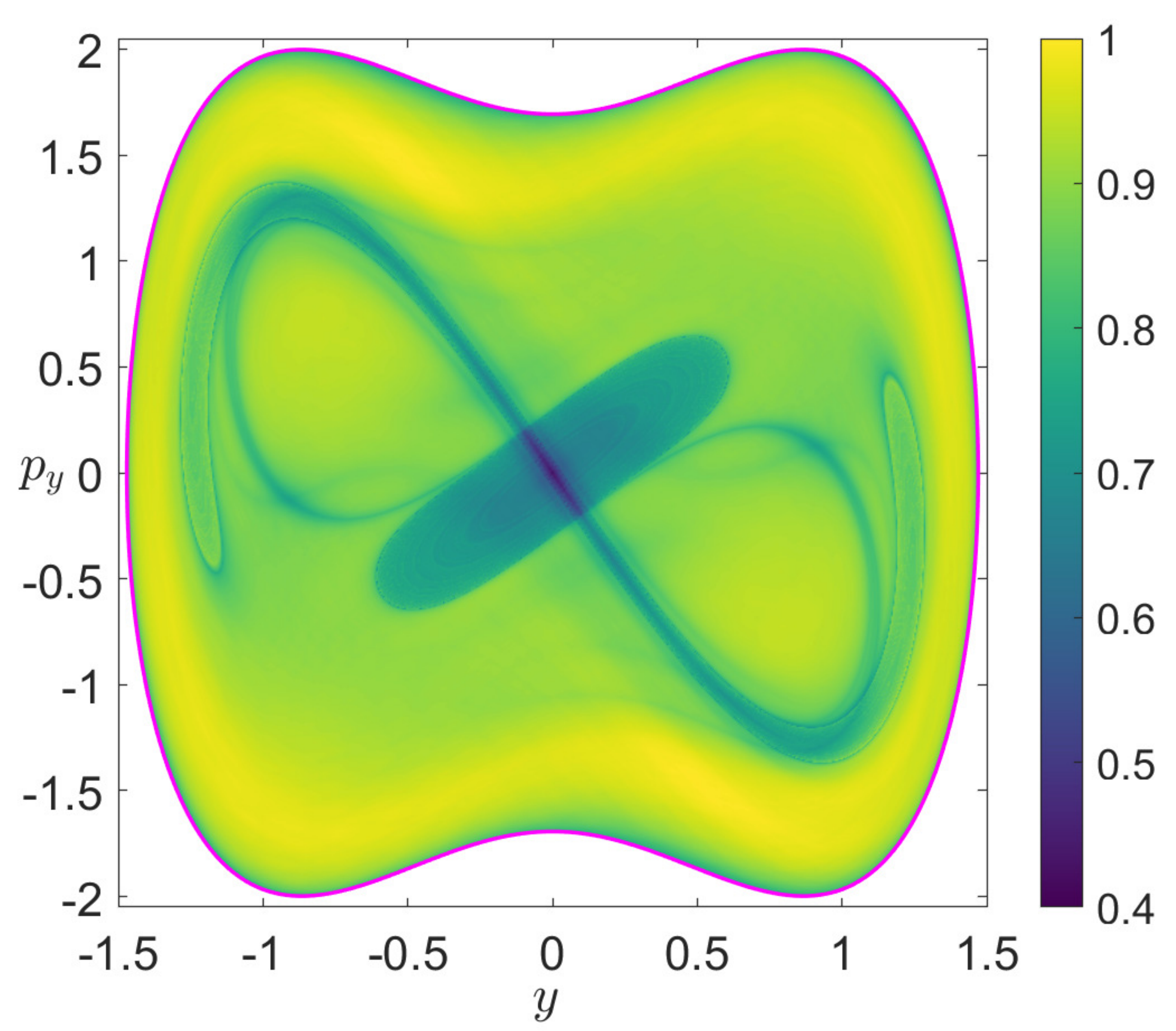}
		B)\includegraphics[scale=0.24]{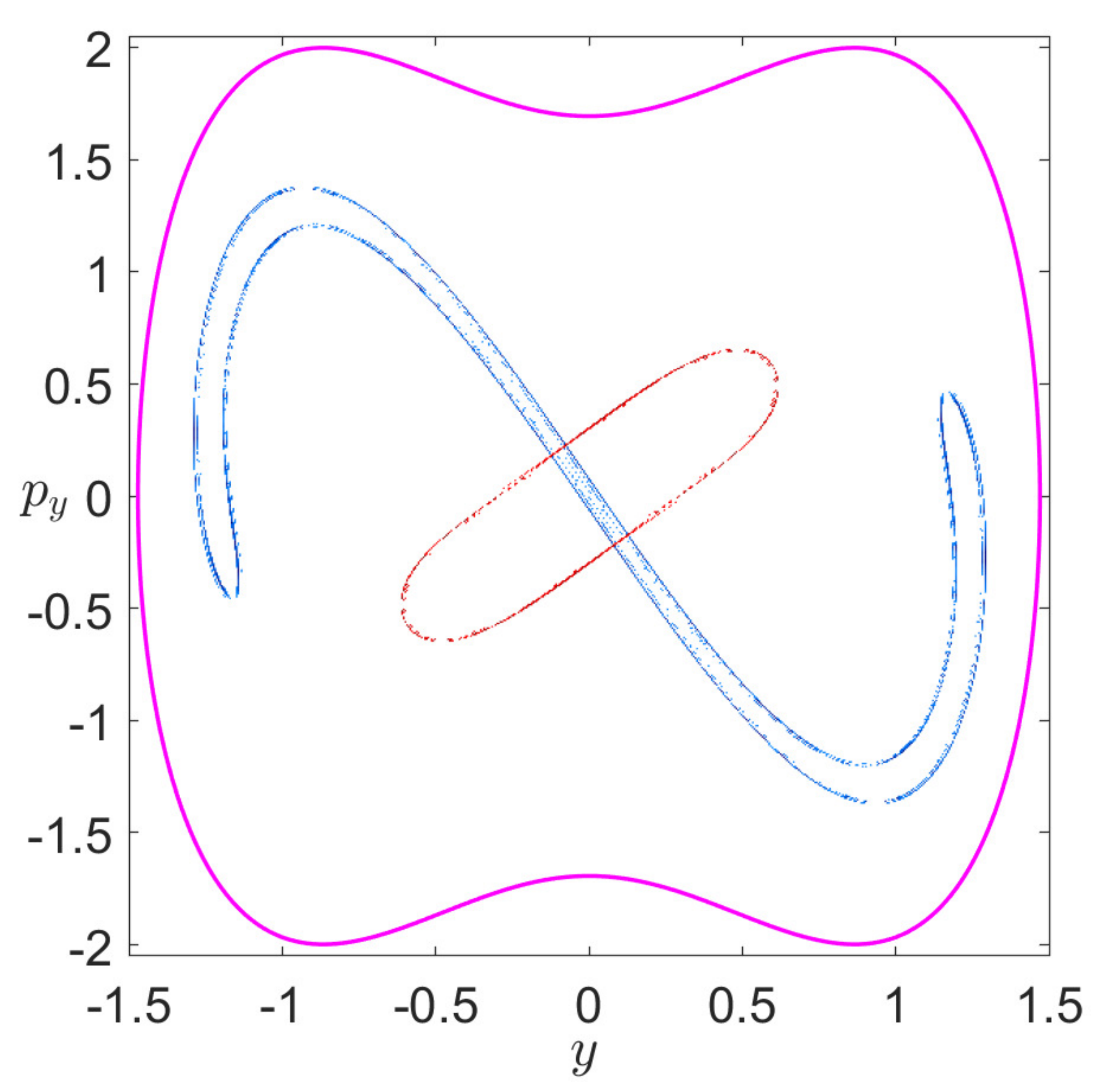}
		C)\includegraphics[scale=0.24]{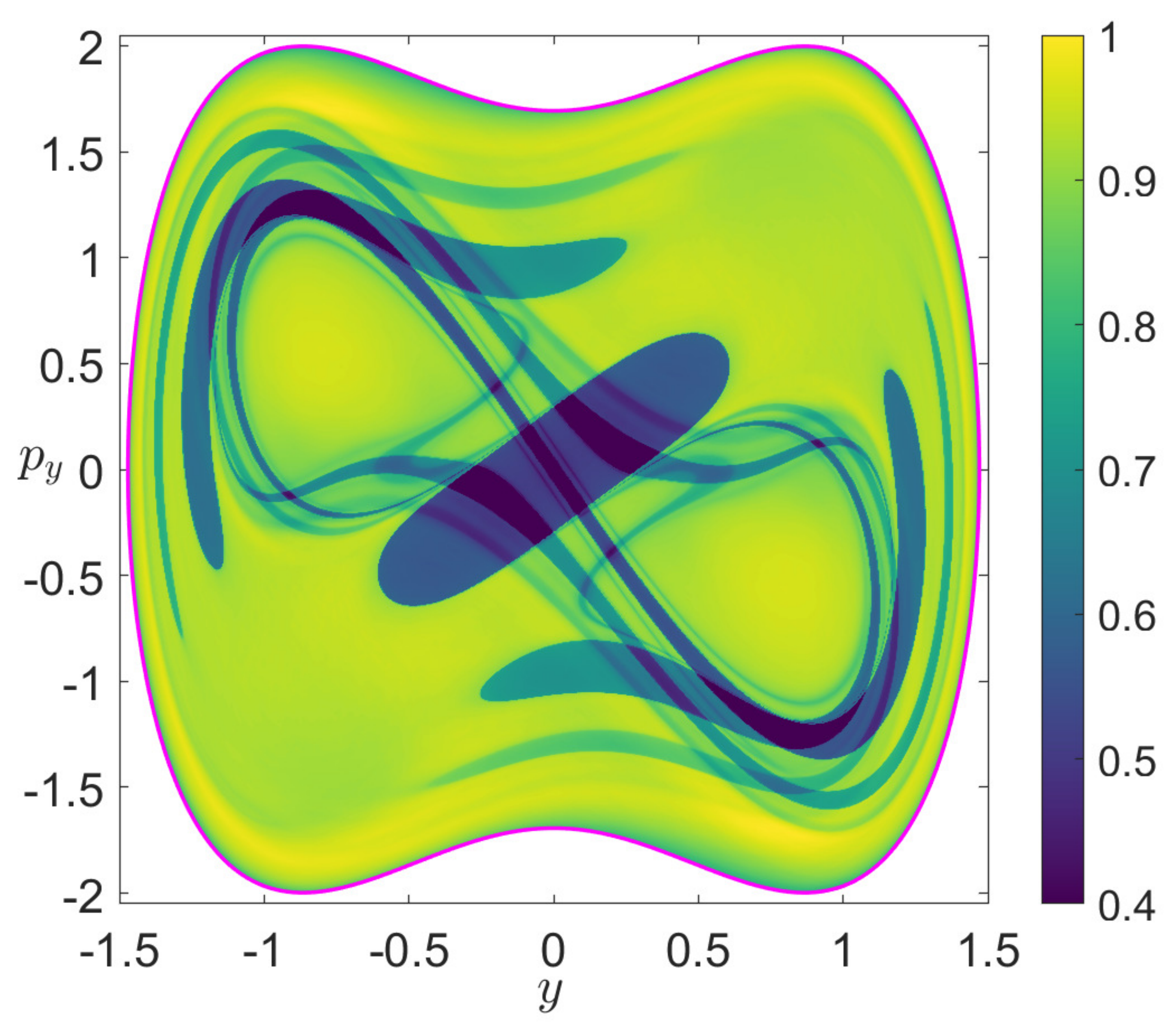}
		D)\includegraphics[scale=0.24]{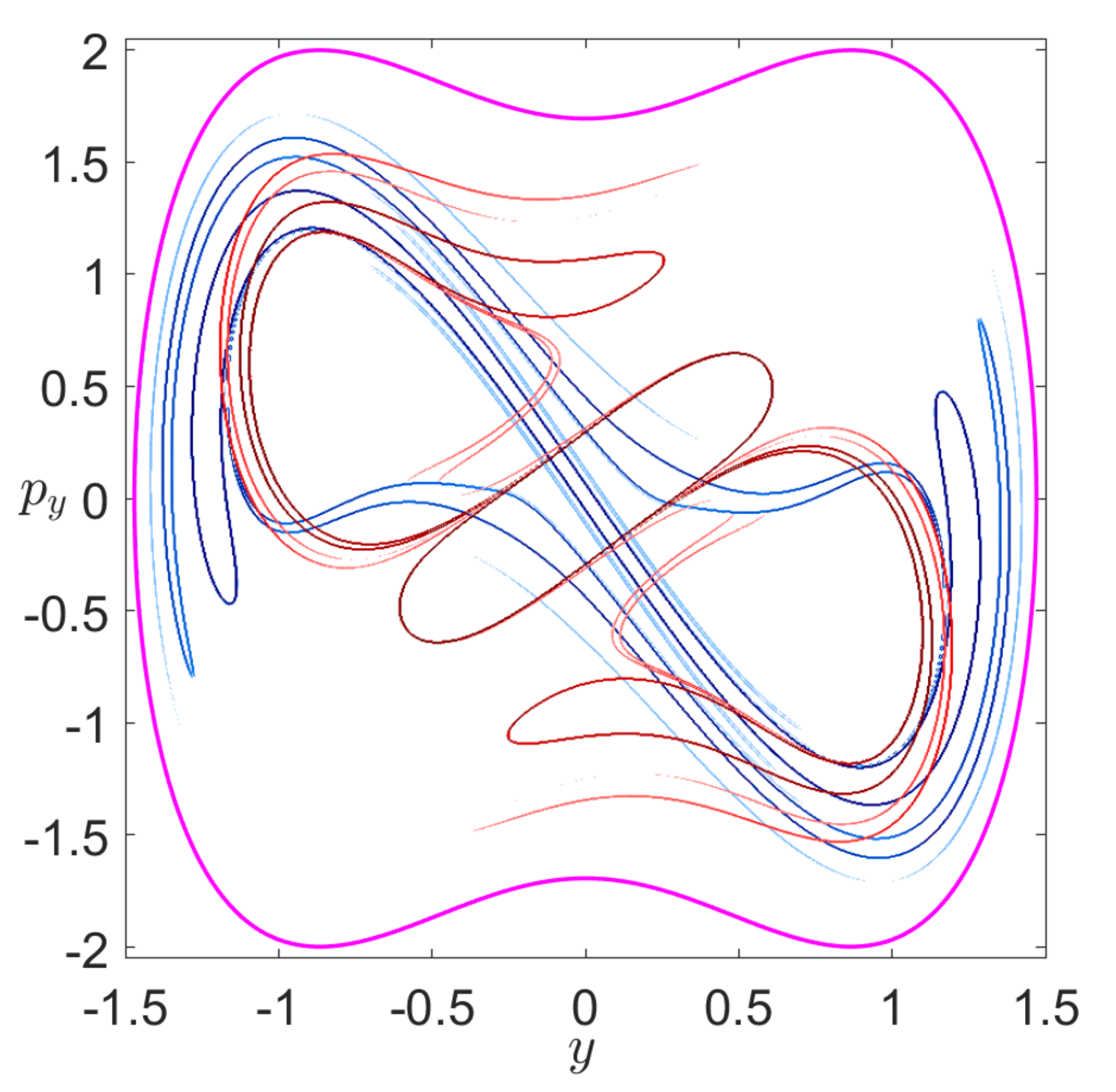}
		E)\includegraphics[scale=0.24]{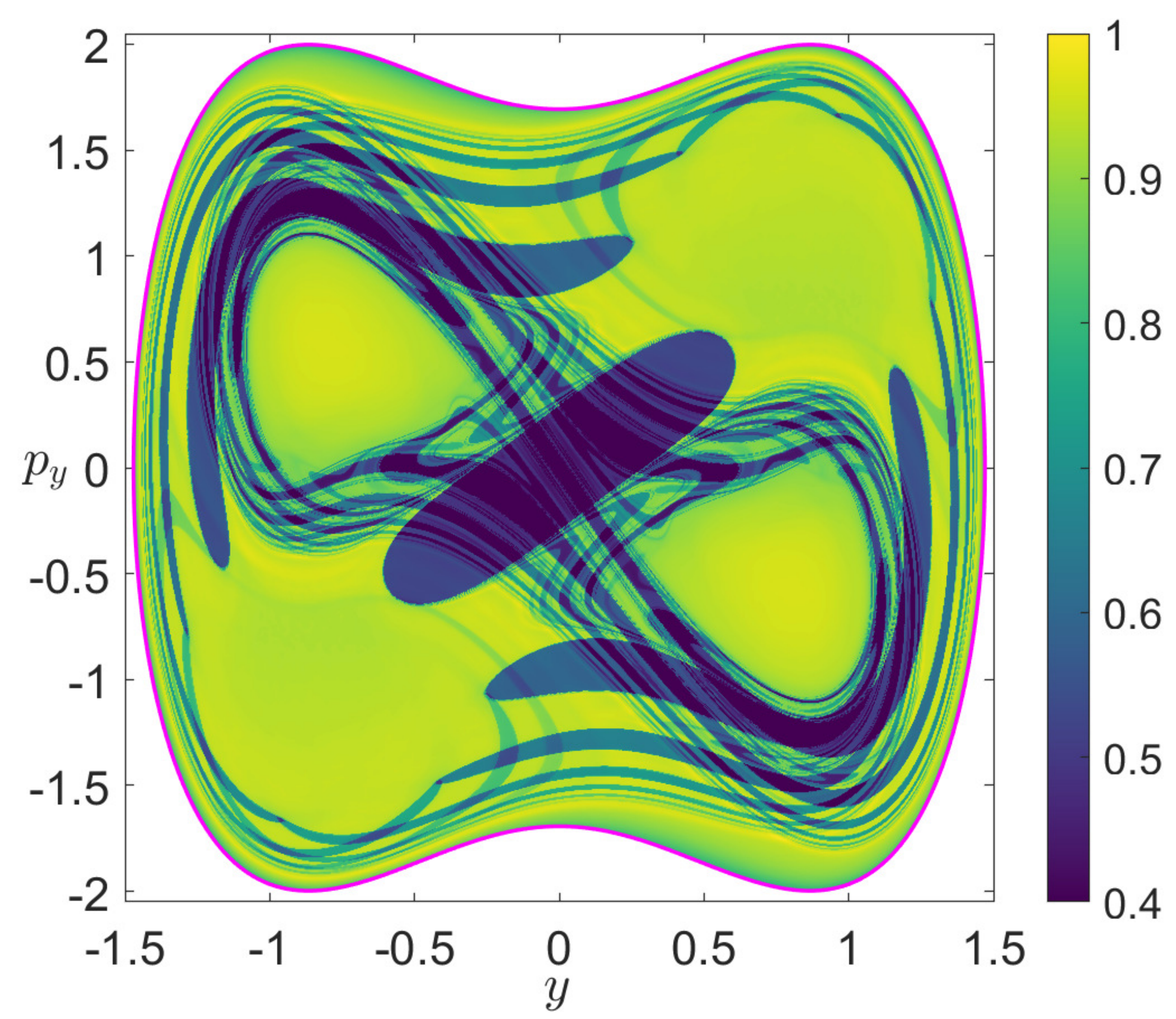}
		F)\includegraphics[scale=0.24]{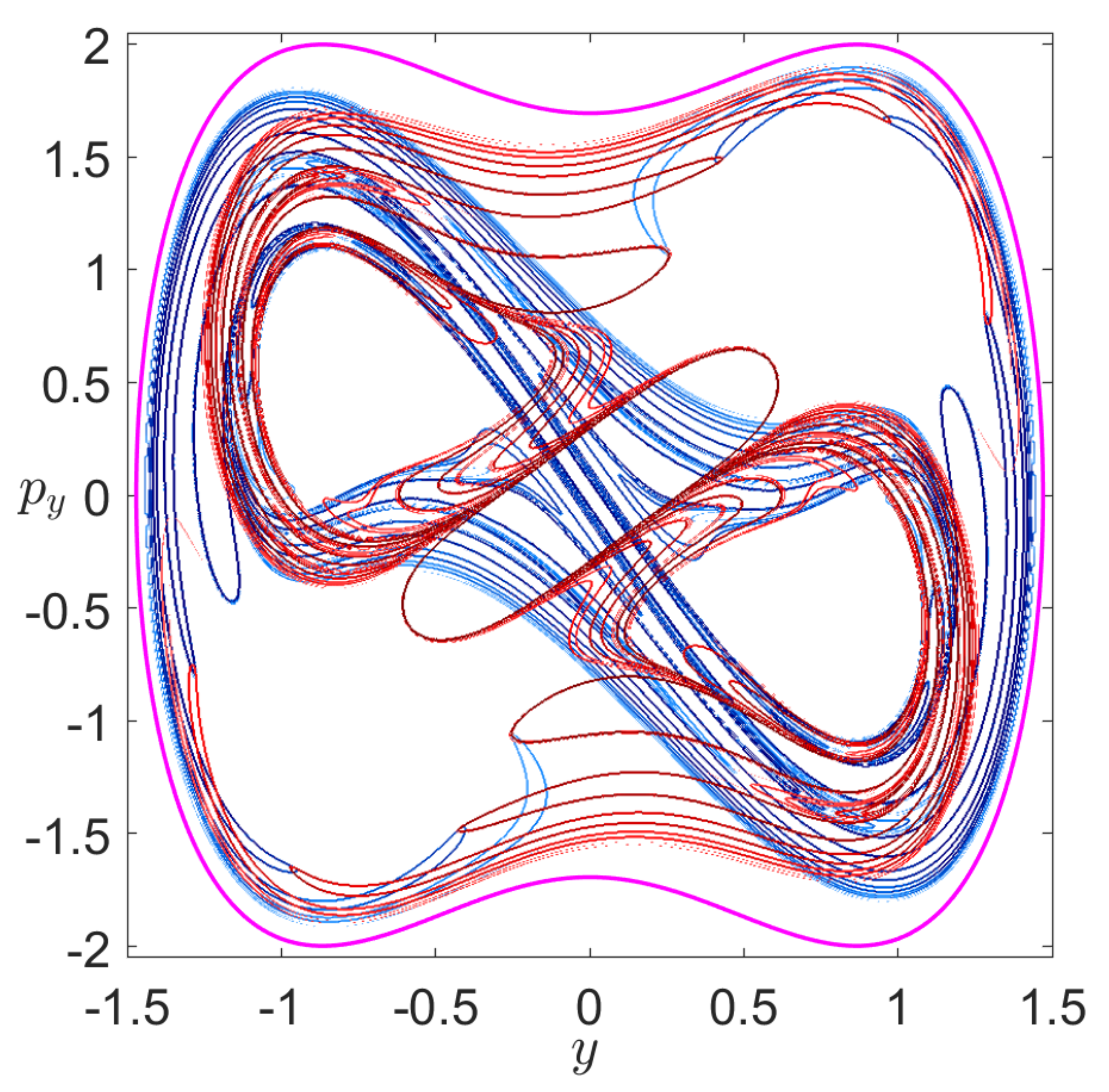}
	\end{center}
	\caption{Lagrangian descriptors calculated on the surface of section $\Sigma_2(H_0)$, where the system's energy is $H_0 = 0.1$, for three different values of the integration time. A) and B) correspond to $\tau = 4$; C) and D) use $\tau = 8$; E) and F) are for $\tau = 16$. In the right column we display the stable (blue) and unstable (red) manifolds extracted from the gradient of the LD scalar field.}
	\label{ld_complexity}
\end{figure}

We finish this appendix by illustrating how LDs can reveal the geometry of invariant manifolds with increasing complexity as the integration time parameter $\tau$ is increased. We do so by calculating LDs on the section $\Sigma_{2}(H_{0})$ for the system with energy $H_{0} = 0.1$, which is, as we have discussed, above that of the index-1 saddle at the origin. We carry out the computation by using the values $\tau = 4,8,16$ and the results obtained are shown in Fig. \ref{ld_complexity}. On the left column we display the scalar field given by the LD diagnostic, and on the right column we demonstrate how the stable (blue) and the unstable (red) manifolds can be extracted from the gradient of the scalar field.

\end{document}